\documentclass[aps,prd,preprint]{revtex4-1}

\usepackage{amsmath}
\usepackage{amssymb}
\usepackage{wasysym}
\usepackage{tikz}
\usetikzlibrary{positioning}
\usepackage[bottom]{footmisc}

\usepackage{mathrsfs}

\usepackage{slashed}
\usepackage{hyperref}

\allowdisplaybreaks

\usepackage{appendix} 

\newcommand{\weight}{\text{\inve}}

\begin{document}

\title{Geometrical Heavy Lifting: Yang-Mills, Spin, and Torsion in Dynamical Projective Gravitation}

\author{Samuel J. Brensinger}
\affiliation{%
 Department of Mathematics \\
 University of Dayton,  OH 45469}
\email{sbrensinger1@udayton.edu}

\author{Patrick Vecera}
\affiliation{%
Department of Mathematics\\
University of California
Santa Barbara, CA 93106-3080}
\email{pvecera@ucsb.edu}

\date{\today}

\begin{abstract}
Thomas-Whitehead (TW) gravity is a gauge theory of gravitation based on projective geometry. The theory maintains projective symmetry through the TW connection, an affine connection over the volume bundle of the spacetime manifold. TW gravity obtains dynamics through Lovelock expansions in the action while preserving general relativity as a weak field limit. In this paper we clarify the process of lifting tensor and spinor fields from spacetime to the volume bundle and demonstrate that a choice of lifting amounts to a gauge fixing condition. This leads to a natural extension of previous work, where we now realize these prior constructions have been restricted to a particular gauge. In pursuit of generality, we also introduce torsion to the TW connection, leading to new dynamics. In particular, the appearance of torsion induces interaction terms involving gravitational coupling with Yang-Mills fields and Dirac spinors. An explicit realization of this is a geometrically sourced chiral mass term arising from the torsion dynamics. 
\end{abstract}

\maketitle

\section{Introduction}

Projective gravity in some form has been around for over a century as a strategy to incorporate reparametrization invariance of geodesics \cite{Veblen1930,Thomas:1925a,Thomas:1925b} in dynamical actions, especially as related to Kaluza-Klein theories. In the context of TW gravity, the covariant derivative on the Thomas cone ties projective geometry to string theory and higher dimensional gravity through projective Gauss-Bonnet terms on the manifold \cite{Brensinger:2020gcv}. This construction allows fields appearing through the TW sector to be interacting and dynamical while preserving general relativity as a weak field limit. 

In the recent work \cite{PalatiniProjective:2017} the authors explored the subtle relationship between the Palatini formalism \cite{palatini} and reparametrization invariance. They explicitly showed that solutions to the connection field equations arising from the Einstein-Hilbert action include more than just the Levi-Civita connection. In particular, the solution space is parametrized by projective transformations of the Levi-Citiva connection, thereby demonstrating a deep relationship between the Palatini formalism and projective geometry. The Palatini formalism of Lovelock gravity has been studied in \cite{Borunda:2008}, wherein the authors examined consistency of solutions to the field equations derived in the Palatini formalism from the Lovelock Lagrangian. In TW gravity the action contains the projective analogue of the Gauss-Bonnet action, which allows a dynamical interpretation of the components of the TW connection. 

Given the fact that TW gravity lives on a bundle over a manifold $M$, it is natural to ask how to make sense of ordinary tensor fields on $M$ in the context of this bundle. A particular answer to this question has been given in the course of the papers \cite{Brensinger:2017gtb,Brensinger:2019mnx,Brensinger:2020gcv}, wherein tensor indices on $M$ can be extended to the Thomas cone by viewing $M$ as a certain section of the bundle. Motivated by a desire to understand the underlying geometry, this paper reexamines the aforementioned procedure through studying the representations of tensor fields on $M$, through which we find that the previously known lift is not unique. Indeed, the possible lifts are parametrized by a specific vector field on $M$. In all previous works this vector has been assumed to vanish identically, a condition dubbed the Roberts gauge after \cite{Roberts, HebdaRobertsExamples}. Furthermore, when considering spinor fields on $M$ such a lift induces a spinor weight, ultimately modifying the fermionic dynamics conscripted by the Dirac equation. In this way there is a natural geometric origin for a chiral mass, as well as other interaction terms involving the TW connection components. 

Thus far all studies of connections in the context of a TW Lagrangian have assumed the absence of torsion \cite{Brensinger:2020gcv}. While at first glance a benign assumption, it is known that including torsion into a gravitational theory can induce gravitational coupling to spinor fields \cite{DiracEquationTorsionZecca} and Yang-Mills gauge fields \cite{SpinAndTorsion}. Including torsion in the TW connection gives rise to several new tensor fields while preserving certain limits that recover general relativity. As such, a study of how torsion modifies the current TW theory will provide theoretical insights and phenomenological implications, especially with regard to gravitational interactions with other fundamental fields. 

Although it is natural to view TW gravity as an extension of general relativity, the historical origin came from 2d quantum gravity \cite{RodgersReview2022}. For the interested reader we include a brief summary of this story - for the full technical details, see \cite{Brensinger:2020gcv}. In 2d quantum gravity, one may integrate the Kirillov two-form over coadjoint orbits \cite{Kirillov:1962,Kirillov:1982kav} of the Kac-Moody and Virasoro algebras \cite{Rai:1989js,Delius:1990pt}, leading to the two-dimensional Wess-Zumino-Witten  \cite{DiVecchia:1984ksr,Witten:1983ar} and Polyakov actions  \cite{Polyakov:1987zb,Polyakov:1981rd}, respectively, which are inherently geometric by construction \cite{Balachandran:1979pc,Balachandran:1986hv,Balachandran:1987st}. In such actions, one promotes coadjoint elements to fields and central extensions to coupling constants. Thus the action interprets a Virasoro coadjoint element as a background field coupling to the Polyakov metric. This element is known in the literature as the diffeomorphism (diff) field \cite{Rai:1989js}, written $\mathcal D_{bc}$, and is analogous to the Yang-Mills connection $A_{\mu}$. The fields $A_{\mu}$ and  $\mathcal{D}_{bc}$ can be rendered dynamical through the Yang-Mills and Thomas-Whitehead (TW) actions \cite{Brensinger:2017gtb,Brensinger:2020gcv,Brensinger:2019mnx}, respectively, resulting in a realization of the coadjoint elements as components of connections in higher dimensions. In \cite{Brensinger:2020gcv} a detailed construction of TW gravity is discussed, including a derivation of the spin connection and Dirac equation on the Thomas cone \cite{Eastwood2}. For a historical overview of the genesis of TW gravity, see \cite{RodgersReview2022}. 

This paper opens with a review of TW gravity in the next section, outlining developments in \cite{Brensinger:2017gtb,Brensinger:2019mnx,Brensinger:2020gcv}. The review is a brief summary of published results, although we formally introduce the Mera convention for index manipulations, which in the literature has been implied but not explicitly defined \cite{MeraVeceraSupermanifolds}. After our review we describe the representation theory behind lifting tensor fields on $M$ to the volume bundle $VM$ and introduce the Roberts gauge, which formally classifies a certain ambiguity in this lifting process that has thus far been unaddressed. Once we have an explicit description of the lifting process, we turn our attention to torsion, describing the TW connection with torsion and constructing associated curvature tensors. We then find the spin connection and build the TW-Dirac action, followed by the next two sections, which are devoted to exploring the interactions between TW gravity and Maxwell/Yang-Mills gauge theories. Finally, in the last section we construct the conglomerate Thomas-Whitehead-Dirac-Yang-Mills action with torsion, encompassing everything in this paper.

\section{Review of Thomas-Whitehead Gravity}

\subsection{Building the Thomas Cone}

Throughout this paper, let $M$ be a smooth orientable Lorentzian manifold. Starting with a (not necessarily metric-compatible) torsionfree affine connection $\Gamma^a_{\ bc}$ on $M$, the fundamental projective invariant $\Pi$ is defined as 
\begin{equation}
\label{eq: Pi definition}
\Pi^a_{\ bc} = \Gamma^a_{\ bc} - \frac{1}{d+1} \big( \Gamma^d_{\ dc} \delta^a_{\ b} + \Gamma^d_{\ db} \delta^a_{\ c} \big),
\end{equation}
which is invariant under projective transformations of the form 
\begin{align}
\Gamma^a_{\ bc} \to \Gamma^a_{\ bc} + \delta^a_{\ b} v_c + \delta^a_{\ c} v_b. 
\end{align}
Letting $J^a_{\ b}$ be the Jacobian of the coordinate transformation $x^a \rightarrow y^a$, the coordinate transformation law of the fundamental projective invariant is 
\begin{align}
\label{eq: Pi transformation}
    \Pi^a_{\ bc} & \to J^a_{\ f} \bigg( \Pi^f_{\ de} \bar{J}^e_{\ c} \bar{J}^d_{\ b} + \frac{\partial^2 x^f}{\partial y^b \partial y^c} \bigg) + \frac{1}{d+1} \big( \bar{J}^m_{\ c} \delta^a_{\ b} + \bar{J}^m_{\ b} \delta^a_{\ c} \big) \frac{\partial}{\partial x^m} \log |J|. 
\end{align}
From equation \ref{eq: Pi transformation} it is apparent that $\Pi$ is not a connection due to the extra terms in the transformation law. To construct a connection realizing projective invariance we follow \cite{Thomas:1925a,Thomas:1925b} and consider instead a connection on the volume bundle $VM$. 

Any smooth function $v:M \rightarrow \mathbb{R}^+$ determines a volume form 
\begin{align}
\label{eq: volume form}
v(x) \, dx^1 \wedge...\wedge dx^d, 
\end{align}
which can be interpreted as a section of the volume bundle $VM$ as in \cite{CrampinSaunders, Thomas:1925a, Thomas:1925b}. The Thomas cone is the bundle $VM$, which is the collection of all such volume sections. The Thomas cone is an $\mathbb{R}^+$ line bundle over $M$, and we use $\lambda$ as the coordinate for the fiber, so the coordinates on $VM$ are $(x^0, x^1,...,x^{d-1}, \lambda)$, where $0 < \lambda < \infty$. Actually, the Thomas cone is well defined even for non-orientable manifolds; for a formal construction, we refer to \cite{Roberts, HebdaRobertsExamples, CrampinSaunders}, where $VM$ is built using the notion of an odd scalar density, amending the definition in equation \ref{eq: volume form} to the case where $M$ is non-orientable. We omit this discussion here only to avoid a lengthy digression into tensor densities. 

Finally, before lifting $\Pi$ to a connection on $VM$ we introduce the Mera convention for Thomas cone coordinates. In the Mera convention we use uppercase letters to denote coordinates on $VM$, while lowercase letters range from $0$ to $d-1$ and denote the corresponding coordinates on $M$, e.g. $x^A = (x^0, \dots, x^a, x^{\lambda})$. This notation collaborates well with tensors of higher rank. Indeed, if $T_{AB}$ is some rank 2 tensor on $VM$, we can express it in block form as 
\begin{equation}
T_{AB} = 
    \left( \begin{array}{@{}c|c@{}}
    T_{ab} & T_{a \lambda} \\ \hline
    T_{\lambda b} & T_{\lambda \lambda}
  \end{array} \right)_{AB} \, ,
\end{equation}
a convention that is immensely beneficial when considering metrics and curvature tensors.

\subsection{Thomas-Whitehead Connection}

The Thomas-Whitehead connection $\tilde{\Gamma}^A_{ \ BC}$ is an affine connection on the Thomas cone that is projectively invariant. The TW connection can be decomposed in components as
\begin{equation}
    \label{eq: Gammatilde}
    \tilde{\Gamma}^{A}_{ \ BC} = \begin{cases}
        \tilde{\Gamma}^a_{\ bc} = \Pi^a_{\ bc} \\
        \tilde{\Gamma}^{\lambda}_{\ bc} = \lambda \mathcal{D}_{bc} \\ 
        \tilde{\Gamma}^a_{\ b \lambda} = \tilde{\Gamma}^a_{\ \lambda b} = \frac{1}{\lambda} \delta^a_{\ b} \\ 
        \tilde{\Gamma}^{\lambda}_{\ b \lambda} = \tilde{\Gamma}^{\lambda}_{\ \lambda b} = \tilde{\Gamma}^{\lambda}_{\ \lambda \lambda} = 0 \\ 
    \end{cases}\, ,
\end{equation}
where $\mathcal{D}_{bc}$ is a (non-tensorial) rank 2 object on $M$ known as the Diffeomorphism (Diff) Field \cite{Rodgers:1994ck}. As $\tilde{\Gamma}$ is a connection, it satisfies 
\begin{align}
\label{eq: TW connection transformation}
\tilde{\Gamma}^A_{\ BC} \to \frac{\partial y^A}{\partial x^D} \frac{\partial x^E}{\partial y^B} \frac{\partial x^F}{\partial y^C} \tilde{\Gamma}^D_{\ EF} + \frac{\partial y^A}{\partial x^D} \frac{\partial^2 x^D}{\partial y^C \partial y^B},    
\end{align}
under coordinate transformations $X^A \to Y^A$ on $VM$, which, using equation \ref{eq: Pi transformation}, implies the transformation law for the Diff field is \cite{Thomas:1925b,Thomas:1925a,Brensinger:2020gcv} 
\begin{align}
\mathcal{D}_{bc} \to \Bigg( \mathcal{D}_{ef} - \frac{\partial}{\partial x^e} j_f + \Pi^d_{\ ef} j_d - j_e j_f \Bigg) \frac{\partial x^e}{\partial y^b} \frac{\partial x^f}{\partial y^c},    
\end{align}
where $j_a = \partial_a \log{J^{-\frac{1}{d+1}}}$. In brief, $\mathcal{D}_{bc}$ lives on $M$ and ensures that $\tilde{\Gamma}$ transforms as a connection over $VM$.

\subsection{Projective geometry}

The projective curvature tensor is built in the usual manner, 
\begin{align}
        \mathcal{K}^A_{\ BCD} = \partial_C \tilde{\Gamma}^A_{\ BD} - \partial_D \tilde{\Gamma}^A_{\ BC} + \tilde{\Gamma}^A_{\ CE} \tilde{\Gamma}^E_{\ BD} - \tilde{\Gamma}^A_{\ DE} \tilde{\Gamma}^E_{\ BC}.
\end{align}
Using equation \ref{eq: Gammatilde} to expand $\tilde{\Gamma}^{A}{}_{BC}$, the only nonvanishing components of $\mathcal{K}^A_{\ BCD}$ are 
\begin{align} 
\label{eq:Projective Curvature Tensor Components} 
\mathcal{K}^{a}_{\ bcd} & = \mathcal{R}^{a}_{\ bcd} + \delta^a_{\ [c}\mathcal{D}_{d]b} \\ 
\mathcal{K}^{\lambda}_{\ bcd} & = \lambda \partial_{[c}\mathcal{D}_{d]b} + \lambda \Pi^{e}_{\ b[d}\mathcal{D}_{c]e}. 
\end{align}
Here $\mathcal{K}^a_{\ bcd}$ is known as the Projective Riemann curvature tensor. The object $\mathcal{K}^{\lambda}_{\ bcd}$ is not itself a tensor on either $M$ or $VM$, but 
\begin{align}
    \mathcal{K}_{bcd} & = \frac{1}{\lambda} \mathcal{K}^{\lambda}_{\ bcd} + g_a \mathcal{K}^a_{\ bcd} 
\end{align}
is tensorial on $M$ and is known as the Projective Cotton-York tensor.

\subsection{The Metric on $VM$}

For any metric $g_{ab}$ on $M$, there is an associated metric on $VM$ given by \cite{Brensinger:2020gcv}
\begin{equation}
\label{eq:Big Metric}
G_{AB} = 
    \left(\begin{array}{@{}c|c@{}}
    g_{ab} + \lambda_0^2 g_a g_b & \frac{\lambda_0^2}{\lambda} g_a \\ \hline
    \frac{\lambda_0^2}{\lambda} g_b & \frac{\lambda_0^2}{\lambda^2}
  \end{array} \right)_{AB}, 
\end{equation}  
where $g_a = -\frac{1}{d+1}\,\partial_a \log \sqrt{|g|}$ with $g$ the metric determinant. It should be noted our choice for $g_{ab}$ is the mostly plus signature, and (regardless of signature) $\lambda$ has the sign of a spacelike component. Our choice for the mostly plus convention is in contrast to previous works including \cite{Brensinger:2017gtb, Brensinger:2020gcv}, which universally use the mostly minus convention. The factor of \(\lambda_0\), like \(\lambda\), has units of length, and ensures that \(G_{AB}\) remains dimensionless. Further, because \(G_{AB}\) depends only on \(g_{ab}\) it is projectively invariant. 

Equation \ref{eq:Big Metric} implies  
\begin{equation} 
\label{eq:Relation Between Metric Determinants} |G| = \frac{\lambda^2_0}{\lambda^2} \; |g|, \qquad  \sqrt{|G|} = \frac{\lambda_0}{\lambda} \sqrt{|g|},  
\end{equation}
where \(g\) is the determinant of \(g_{ab}\) on \(M\), a relationship that is useful when constructing actions as it provides a natural measure over $VM$. Lastly, the inverse of \(G_{AB}\) is given by
\begin{equation}
\label{eq:Big Metric Inverse}
  G^{AB} = 
  \left(\begin{array}{@{}c|c@{}}
    g^{ab} & - \lambda g^{am} g_m \\ \hline
    - \lambda g^{bm}g_m & \frac{\lambda^2}{\lambda_0^2} + \lambda^2 g^{mn} g_m g_n 
  \end{array} \right)^{AB} 
\end{equation}
where \(g^{ab}\) is the inverse of the spacetime metric \(g_{ab}\).

\section{Vectors on the Thomas Cone}

A general procedure for lifting a vector from $M$ to $VM$ was given in \cite{Brensinger:2020gcv}. Given a vector $V^a$ on $M$, it has a transformation law 
\begin{align}
    V^a \to J^a_{\ b} V^b,
\end{align}
under a change of coordinates $x^a \to y^a$ on $M$. The simple extension $V^A = (V^a, 0)$ is not tensorial under the transformation $X^A \to Y^A$ on $VM$, but the conglomerate 
\begin{align}
\label{eq: VectorMtoVM}
    V^A = (V^a, - \lambda V^b \kappa_b )
\end{align}
is. Here $\kappa_b$ is any object that has the transformation law 
\begin{align}
\label{eq: KappaTransformation}
        \kappa_a \to \bar{J}^b_{\ a} \kappa_b + \bar{J}^b_{\ a} j_b, 
\end{align} 
under a transformation $x^a \to y^a$ on $M$. This is identical to the transformation law for $g_a$, and is the origin for the factors of $g_a$ in the metric on $VM$ in equation \ref{eq:Big Metric}. Similar to vectors, a one-form $X_a$ can be extended to $VM$ by taking 
\begin{align}
\label{eq: One-form from M to VM}
X_A = (X_a + \kappa_a, \tfrac{1}{\lambda} ), 
\end{align}
and the combination of equations \ref{eq: VectorMtoVM} and \ref{eq: One-form from M to VM} preserves the inner product on $VM$, i.e. $V^A X_A = V^a X_a$. 

One may wonder if the converse is true; in other words, if any vector and one-form on $VM$ are of the form in equations \ref{eq: VectorMtoVM} and \ref{eq: One-form from M to VM}, respectively. To answer this, suppose we have some one-form $X_A$ on $VM$ which  transforms by 
\begin{align}
    X_A \to \bar{J}^B_{\ A} X_B. 
\end{align}
This expands as  
\begin{align}
    X_A \to \bar{J}^b_{\ A} X_b + \bar{J}^{\lambda}_{\ A} X_{\lambda}.
\end{align}
or, in components, 
\begin{align}
    ( X_a, X_{\lambda} ) \to ( \bar{J}^b_{\ a} X_b + \bar{J}^{\lambda}_{\ a} X_{\lambda}, \bar{J}^b_{\ \lambda} X_b + \bar{J}^{\lambda}_{\ \lambda} X_{\lambda} ). 
\end{align}
Evaluating this rule with the coefficients from \ref{eq: Jacobian on VM}, we find 
\begin{align}
    ( X_a, X_{\lambda} ) \to ( \bar{J}^b_{\ a} X_b -\bar{J}^{b}_{\ a} j_b X_{\lambda}, |J|^{\tfrac{1}{d+1}} X_{\lambda} ), 
\end{align}
which exactly fixes $X_A$ to be of the form in equation \ref{eq: One-form from M to VM}. 

To classify vectors on $VM$, we can work in the other direction; in the review article \cite{CrampinSaunders}, as well as \cite{Brensinger:2017gtb}, it is shown that the Lie algebra structure in projective geometry requires the inner product to be invariant under lifting to $VM$, establishing the relation $V^A X_A = V^a X_a$ prior to introduction of geometric structures. Substituting the coefficients from equation \ref{eq: One-form from M to VM} into the inner product relation, a brief computation reveals that any vector on $VM$ must take the form in equation \ref{eq: VectorMtoVM}. 

Actually, this is not quite true; as observed in \cite{Brensinger:2020gcv}, if we have a vector $V^a$ on $M$, then for any $V_{\perp}$ that is a smooth $\mathbb{R}^+$-valued function over $M$, both 
\begin{align}
(V^a, - \lambda V^b \kappa_b ) \quad \text{and} \quad (V^a, - \lambda V^b \kappa_b + \lambda V_{\perp} )
\end{align}
project to $V^a$. This should come as no surprise, as such a choice of $V_{\perp}$ is perpendicular to $M$, so including this contribution in a vector amounts to moving to a different section of $VM$ over $M$. Indeed, we can remove this term by an appropriate choice of projective transformation, by redefining $\kappa_b$ to take $V_{\perp}$ into account, or by rescaling $\lambda$. Thus given the plethora of means for expunging $V_{\perp}$ we will not be unduly concerned with perpendicular components. 

To sum up, for any particular choice of $\kappa_b$ and $V_{\perp}$, there is an exact correspondence between vectors on $M$ and vectors on $VM$ given in equation \ref{eq: VectorMtoVM}. As such we will follow the conventions in \cite{Brensinger:2020gcv} and assume $V_{\perp} = 0$ and $\kappa_b = g_b$, leaving discussion of these degrees of freedom to future research. This condition is known as the Roberts gauge \cite{Roberts,HebdaRobertsExamples}, and will be frequently assumed in this work. 

At this point the above demonstration may reasonably seem unapologetically esoteric. However, as we will see, the correspondence between vector fields on $M$ and $VM$ has important consequences with regard to gauge theories on $VM$, and therefore really warrants a close inspection. 

Before concluding our review we describe one final object that appears naturally and simplifies numerous computations. Above we observed that $g_a$ has a nontrivial transformation law. From equation \ref{eq: One-form from M to VM} we see that the extension $g_A = (g_a, \tfrac{1}{\lambda})$ is tensorial on $VM$ and is actually the lift of the zero covector from $M$ to $VM$. This is why $g_A$, not just $g_a$, is what appears in $G_{AB}$ if $\lambda$ components are included.

\section{Torsionful Thomas-Whitehead Connection}

\subsection{General Projective Transformations}

As discussed in the introduction, as well as in  \cite{Brensinger:2017gtb,Brensinger:2019mnx,Brensinger:2020gcv}, the Thomas-Whitehead connection has been constructed to ensure that connections on $M$ related by a transformation of the form 
\begin{align}
    \Gamma^a_{\ bc} \to \Gamma^a_{\ bc} + \delta^a_{\ b} v_c + \delta^a_{\ c} v_b  
\end{align}
are contained in the same equivalence class and produce identical physics. However, the more general transformation 
\begin{align}
\label{eq: lopsided projective transformation}
    \Gamma^a_{\ bc} \to \Gamma^a_{\ bc} + \delta^a_{\ b} v_c + \delta^a_{\ c} w_b,  
\end{align}
where generally $w_c \neq v_c$, still preserves equivalence classes of geodesics. Therefore, a true gauge theory of projective gravity really should be broad enough to include these `lopsided' transformations. As we will see, constructing $\Pi$ in such a manner as to be invariant under transformations such as in equation \ref{eq: lopsided projective transformation} naturally introduces torsion into the TW connection.

\subsection{Definition of $\Pi$}

Constructing $\Pi$ to be invariant under symmetric projective transformations was accomplished by starting with $\Gamma$ and subtracting off the trace. Generalizing the fundamental projective invariant to the asymmetric case can be accomplished in a similar manner, but where we must now take into account the asymmetry of $\Gamma$. This can be achieved by defining $\Pi$ to be 
\begin{align}
\label{eq: Pi Torsion Definition}
    \Pi^a_{\ bc} & = \Gamma^a_{\ bc} - \frac{1}{d+1} \bigg( \frac{d}{d-1} \Gamma^e_{\ ec} - \frac{1}{d-1} \Gamma^e_{\ ce} \bigg) \delta^a_{\ b} - \frac{1}{d+1} \bigg( \frac{d}{d-1} \Gamma^e_{\ be} - \frac{1}{d-1} \Gamma^e_{\ eb} \bigg) \delta^a_{\ c},   
\end{align}
which is invariant under both $\Gamma^a_{\ bc} \to \Gamma^a_{\ bc} + \delta^a_{\ b} v_c$ and $\Gamma^a_{\ bc} \to \Gamma^a_{\ bc} + \delta^a_{\ c} w_b$. The symmetric part of $\Pi$ is given by 
\begin{align}
    \frac{1}{2} \Pi^a_{\ (bc)} & = \frac{1}{2} \bigg( \Gamma^a_{\ bc} - \frac{1}{d+1} \bigg( \Gamma^e_{\ ec} + \Gamma^e_{\ ce} \bigg) \delta^a_{\ b} + \Gamma^a_{\ cb} - \frac{1}{d+1} \bigg( \Gamma^e_{\ eb} + \Gamma^e_{\ be} \bigg) \delta^a_{\ c} \bigg),  
\end{align}
(note the factor of $\tfrac{1}{2}$!) which reduces to the definition in equation \ref{eq: Pi definition} when the connection is symmetric. By nature of equation \ref{eq: Pi Torsion Definition}, $\Pi$ now has an antisymmetric part given by
\begin{align}
    \frac{1}{2} \Pi^a_{\ [bc]} & = \frac{1}{2} \bigg( \Gamma^a_{\ bc} - \Gamma^a_{\ cb} - \frac{1}{d-1} \bigg( \Gamma^e_{\ ec} - \Gamma^e_{\ ce} \bigg) \delta^a_{\ b} + \frac{1}{d-1} \bigg( \Gamma^e_{\ eb} - \Gamma^e_{\ be} \bigg) \delta^a_{\ c} \bigg), 
\end{align}
which can be identified with the canonical torsion tensor with the trace removed.  

The coordinate transformation law for an arbitrary (noncompatible, torsionful) affine connection under the transformation $x^a \to y^a$ is just the usual law 
\begin{align}
    \Gamma^a_{\ bc} \to J^a_{\ d} \left( \bar{J}^e_{\ b} \bar{J}^f_{\ c} \Gamma^d_{\ ef} + \frac{\partial^2 x^d}{\partial y^b \partial y^c} \right), 
\end{align}
which, using equation \ref{eq: Trace of Gamma transformation}, descends to the transformation of $\Pi$ as
\begin{align}
    \Pi^a_{\ bc} & \to  J^a_{\ d} \left( \bar{J}^e_{\ b} \bar{J}^f_{\ c} \Pi^d_{\ ef} + \frac{\partial^2 x^d}{\partial y^b \partial y^c} \right) + \frac{1}{d+1} \left( \delta^a_{\ b} \bar{J}^m_{\ \ c} + \delta^a_{\ c} \bar{J}^m_{\ \ b} \right) \partial_m \log |J|. 
\end{align}
Because this is the same transformation rule as in the torsionfree case, we can use $\Pi$ to build $\tilde{\Gamma}$ as before. However, because $\Pi$ is now allowed to contain antisymmetric degrees of freedom, the more general TW connection will naturally contain torsion on $VM$.

\subsection{TW Connection with Torsion}

Before presenting the components of $\tilde{\Gamma}$, a brief digression on conventions is in order. Let $\hat{\Gamma}^a_{\ bc}$ denote the Levi-Civita connection on $M$, while $\Gamma^a_{\ bc}$ is some other arbitrary affine connection. Because the difference of two affine connections is tensorial, we can write 
\begin{align}
    \Gamma^a_{\ bc} - \hat{\Gamma}^a_{\ bc} = C^a_{\ bc} + T^a_{\ bc}, 
\end{align}
where $T^a_{\ bc} = -T^a_{\ cb}$ is the torsion tensor \cite{SpinAndTorsion} and $C^a_{\ bc} = C^a_{\ cb}$ is the Palatini field \cite{Brensinger:2020gcv}. This provides a useful formulation as now we are able to express all quantities in terms of tensors and the Levi-Civita connection. These fields will appear in the connection only with their traces removed, i.e. 
\begin{align}
    \hat{\Pi}^a_{\ bc} & = \hat{\Gamma}^a_{\ bc} - \tfrac{1}{d+1} \delta^a_{\ b} \hat{\Gamma}^e_{\ ec} - \tfrac{1}{d+1} \delta^a_{\ c} \hat{\Gamma}^e_{\ eb}, \nonumber \\ 
    \overline{C}^a_{\ bc} & = C^a_{\ bc} - \tfrac{1}{d+1} \delta^a_{\ b} C^e_{\ ec} - \tfrac{1}{d+1} \delta^a_{\ c} C^e_{\ eb}, \nonumber \\ 
    \overline{T}^a_{\ bc} & = T^a_{\ bc} - \tfrac{1}{d+1} \delta^a_{\ b} T^e_{\ ec} - \tfrac{1}{d+1} \delta^a_{\ c} T^e_{\ eb}.  
\end{align}
The TW connection then takes the form 
\begin{equation}
\label{eq: TW Connection with torsion}
    \tilde{\Gamma}^{A}_{ \ BC} = \begin{cases}
        \tilde{\Gamma}^a_{\ bc} = \Pi^a_{\ bc} = \hat{\Pi}^a_{\ bc} + \overline{C}^a_{\ bc} + \overline{T}^a_{\ bc} \\
        \tilde{\Gamma}^{\lambda}_{\ bc} = \lambda \mathcal{D}_{bc} + \lambda \mathcal{V}_{bc} \\ 
        \tilde{\Gamma}^a_{\ \lambda b} = \frac{1}{\lambda} ( \delta^a_{\ b} + \mathfrak{u}^a_{\ b} ) \\ 
        \tilde{\Gamma}^a_{\ b \lambda} = \frac{1}{\lambda} ( \delta^a_{\ b} - \mathfrak{u}^a_{\ b} ) \\ 
        \tilde{\Gamma}^{\lambda}_{\  \lambda b} = - \tilde{\Gamma}^{\lambda}_{\ b \lambda} =  \mathfrak{a}_b \\ 
        \tilde{\Gamma}^{\lambda}_{\  \lambda \lambda} = \tilde{\Gamma}^a_{\ \lambda \lambda} = 0 \\ 
    \end{cases}\, .
\end{equation}
Here $\mathcal{D}_{bc}$ is unchanged from equation \ref{eq: Gammatilde}, but $\mathcal{V}_{bc}$ can be any antisymmetric rank two tensor, which we dub the $\textit{Veblen 2-form}$ after \cite{Veblen1930}. The field $\mathfrak{u}^a_{\ b}$ is an arbitrary rank two symmetric tensor (i.e. $\mathfrak{u}_{ab} = \mathfrak{u}_{ba}$), while $\mathfrak{a}_b$ can be any rank one tensor. We collectively refer to the fields $\overline{T}, \mathcal{V}, \mathfrak{u}$, and $\mathfrak{a}$ as the $\textit{torsion fields}$ as they are the constituent components of torsion on $VM$. It may be worth remarking that $\overline{T}$ is the tracefree part of the usual torsion tensor, such as in Einstein-Cartan gravity \cite{SpinAndTorsion}, but to the best of our knowledge there is no analogue for the other torsion fields. Likewise, the role of torsion and the Palatini field have been explored in \cite{MetricAffineThesis}, although not in the context of TW gravity. 

Examining the transformation law for $\mathcal{D}$, we see that using the coefficients from equation \ref{eq: TW Connection with torsion}, as well as the transformation rules in \ref{eq: Jacobian on VM}, we get
\begin{align}
    \tilde{\Gamma}^{\lambda}_{\ bc} & \to \lambda |J|^{- \frac{1}{d+1}} \bigg( j_d \Pi^d_{\ ef} + \mathcal{D}_{ef} - j_f j_e - \partial_f j_e \bigg) \bar{J}^e_{\ b} \bar{J}^f_{\ c}, 
\end{align}
which is identical to the torsionfree case, meaning that adding $\mathcal{V}_{bc}$ to this component of $\tilde{\Gamma}$ does not change the properties of $\mathcal{D}_{bc}$.

\section{Projective Riemann Curvature Tensor}

Because we now have asymmetric connection coefficients, we must be careful about distinguishing conventions for covariant derivatives. We choose our convention where the derivative index on covariant derivative operators should always be the $\textit{first}$ index in $\Gamma$ and $\tilde{\Gamma}$, so for example 
\begin{align}
    \tilde{\nabla}_A V^B = \partial_A V^B + \tilde{\Gamma}^B_{\ AE} V^E, & \qquad \tilde{\nabla}_A V_B = \partial_A V_B - \tilde{\Gamma}^E_{\ AB} V_E, \nonumber \\ 
    \nabla_a V^b = \partial_a V^b + \Gamma^b_{\ ae} V^e, & \qquad \nabla_a V_b = \partial_a V_b - \Gamma^e_{\ ab} V_e. 
\end{align}

We are now ready to define the curvature tensor. Usually curvature tensors are commutators of covariant derivatives. However, for a gravitational theory built from a general affine connection the definition of curvature should contain a term proportional to the torsion tensor, as discussed in the timeless text \cite{TensorsDifferentialFormsBookLovelockRund}. For a more recent understanding of the role of this construction in metric-affine models of gravity, see \cite{ModifiedGravityandCosmologyBook}. In our context the curvature tensor is 
\begin{align}
\label{eq: Kurvature with Kommutator of Kovariant Derivatives}
    \mathcal{K}^A_{\ BCD} V^B & = [\tilde{\nabla}_C, \tilde{\nabla}_D] V^A + ( \tilde{\Gamma}^B_{\ CD} - \tilde{\Gamma}^B_{\ DC} ) \tilde{\nabla}_B V^A, 
\end{align}
a detailed derivation of which is in the appendix. Equation \ref{eq: Kurvature with Kommutator of Kovariant Derivatives} expands to 
\begin{align}
    \mathcal{K}^A_{\ BCD} & = \partial_C \tilde{\Gamma}^A_{\ DB} - \partial_D \tilde{\Gamma}^A_{\ CB} + \tilde{\Gamma}^A_{\ CE} \tilde{\Gamma}^E_{\ DB} -  \tilde{\Gamma}^A_{\ DE} \tilde{\Gamma}^E_{\ CB}. 
\end{align}
Using the components of the TW connection from equation \ref{eq: TW Connection with torsion}, the entries of $\mathcal{K}^A_{\ BCD}$ are 
\begin{align}
    \mathcal{K}^{a}_{\ bcd} & =  \mathcal{R}^a_{\ bcd} + ( \delta^a_{\ c} - \mathfrak{u}^a_{\ c} ) \big( \mathcal{D}_{db} + \mathcal{V}_{db} \big) - ( \delta^a_{\ d} - \mathfrak{u}^a_{\ d} ) \big( \mathcal{D}_{cb} + \mathcal{V}_{cb} \big) \nonumber \\ 
    \mathcal{K}^{a}_{\ bc \lambda} & = - \mathcal{K}^{a}_{\ b \lambda c} = \tfrac{1}{\lambda} \big( \partial_c \mathfrak{u}^a_{\ b} + \Pi^a_{\ ce} \mathfrak{u}^e_{\ b} - \mathfrak{u}^a_{\ e} \Pi^e_{\ cb} + ( \delta^a_{\ c} - \mathfrak{u}^a_{\ c} ) \mathfrak{a}_b \big), \nonumber \\ 
    \mathcal{K}^{a}_{\ \lambda cd} & = \tfrac{1}{\lambda} \big( \partial_{[d} \mathfrak{u}^a_{\ c]} + \overline{T}^a_{\ cd} - \Pi^a_{\ [c|e|} \mathfrak{u}^e_{\ d]} + \delta^a_{\ [d} \mathfrak{a}_{c]} - \mathfrak{u}^a_{\ [d} \mathfrak{a}_{c]} \big), \nonumber \\ 
        \mathcal{K}^{a}_{\ \lambda c \lambda} & = - \mathcal{K}^{a}_{\ \lambda \lambda c} = \tfrac{1}{\lambda^2} ( \mathfrak{u}^a_{\ e} \mathfrak{u}^e_{\ c} - \mathfrak{u}^a_{\ c} ), \nonumber \\ 
    \mathcal{K}^{\lambda}_{\ bcd} & = \lambda \big( \partial_{[c} \mathcal{D}_{d]b} + \partial_{[c} \mathcal{V}_{d]b} + \Pi^e_{\ [d|b|} \mathcal{D}_{c]e} + \Pi^e_{\ [d|b|} \mathcal{V}_{c]e} + \mathfrak{a}_{[d} \mathcal{D}_{c]b} + \mathfrak{a}_{[d} \mathcal{V}_{c]b} \big), \nonumber \\ 
        \mathcal{K}^{\lambda}_{\ bc \lambda} & = - \mathcal{K}^{\lambda}_{\ b \lambda c} = \partial_c \mathfrak{a}_b + ( \mathcal{D}_{ce} + \mathcal{V}_{ce} ) \mathfrak{u}^e_{\ b} - \mathfrak{a}_c \mathfrak{a}_b -  \mathfrak{a}_e \Pi^e_{\ cb}, \nonumber \\ 
    \mathcal{K}^{\lambda}_{\ \lambda cd} & = \partial_{[d} \mathfrak{a}_{c]} + \mathfrak{u}^e_{\ [c} \mathcal{D}_{d]e} + \mathfrak{u}^e_{\ [c} \mathcal{V}_{d]e} + 2 \mathcal{V}_{cd}, \nonumber \\ 
        \mathcal{K}^{\lambda}_{\ \lambda c \lambda} & = - \mathcal{K}^{\lambda}_{\ \lambda \lambda c} = \tfrac{1}{\lambda} ( \mathfrak{a}_e \mathfrak{u}^e_{\ c} - \mathfrak{a}_c ), \nonumber \\ 
    \mathcal{K}^{a}_{\ b \lambda \lambda} & =  \mathcal{K}^{a}_{\ \lambda \lambda \lambda} =  \mathcal{K}^{\lambda}_{\ b \lambda \lambda} = \mathcal{K}^{\lambda}_{\ \lambda \lambda \lambda} = 0. 
\end{align}
Here the equi-projective Riemann tensor is 
\begin{align}
    \mathcal{R}^a_{\ bcd} & = \partial_c \Pi^a_{\ db} - \partial_d \Pi^a_{\ cb} + \Pi^a_{\ ce} \Pi^e_{\ db} -  \Pi^a_{\ de} \Pi^e_{\ cb},  
\end{align}
which is projectively invariant but not tensorial. When the torsion fields vanish the only surviving components of $\mathcal{K}^A_{\ BCD}$ are $\mathcal{K}^a_{\ bcd}$ and $\mathcal{K}^{\lambda}_{\ bcd}$, recovering the components in \cite{Brensinger:2020gcv} as the limiting case.

\section{Spinor Fields and the Dirac Equation}

\subsection{Setting up Spinors: Frame Fields and Gamma Matrices}

Before constructing objects required for studying spinor fields, we begin by fixing conventions. In addition to the assumption of orientability throughout, in this section we assume $M$ (at least locally) admits spinors. Further, we impose the condition $M$ is even dimensional so that the dimension of the spin representation on $M$ is the same as that on $VM$. For index placement we continue to follow the Mera convention but also introduce underlined indices to indicate flattened coordinates, e.g. $V^A$ is a vector on $VM$, but $V^{\underline{A}}$ is a vector in $d+1$-dimensional Minkowski space. Much of the setup involved here is fairly standard material recast in the setting of projective geometry, a helpful resource for which is \cite{SpinorStuffPoplawski}, especially regarding spinors on a curved spacetime. 

In order to study spinors on $VM$ it is necessary to construct tetrads. Because $G_{AB}$ is not changed from equation \ref{eq:Big Metric} by including torsion, we construct frame fields using equations \ref{eq:Big Metric} and \ref{eq:Big Metric Inverse}, so tetrads on $VM$ take the form 
\begin{equation}
\label{eq: frame fields on VM}
  \tilde{e}^M_{\ \, \underline{A}} = 
  \left(\begin{array}{@{}c|c@{}}
    e^m_{\ \underline{a}} & 0 \\\hline
    - \lambda e^p_{\ \underline{a}} g_p & \frac{\lambda}{\lambda_0} 
  \end{array} \right)^M_{\ \, \underline{A}}, \qquad \quad \tilde{e}_{\ \ M}^{\underline{A}} = 
  \left(\begin{array}{@{}c|c@{}}
    e_{\ m}^{\underline{a}} & 0 \\ \hline
    \lambda_0 g_m & \frac{\lambda_0}{\lambda}
  \end{array} \right)^{\underline{A}}_{\ \ M} \, ,
\end{equation}
where $e^m_{\ \underline{a}}$ and $e_{\ m}^{\underline{a}}$ are the frame fields for $g_{ab}$ on $M$. With tetrads in place, we can move between spacetime and flattened coordinates on both $VM$ and $M$, e.g. $V^A = \tilde{e}^A_{\ \underline{A}} V^{\underline{A}}$ while $V^a = e^a_{\ \underline{a}} V^{\underline{a}}$. 

In addition to tetrads, spinor computations generally require gamma matrices. Here gamma matrices come in two forms, those on $M$ and those on $VM$. The gamma matrices on $M$ satisfy the usual relation
\begin{align}
    \{ \gamma^m, \gamma^n \} = - 2 g^{mn} I_\mathfrak{D},
\end{align}
where the minus sign is for consistency with our signature of metric, $\mathfrak{D} = 2^{\lfloor \frac{d}{2} \rfloor }$ is the dimension of the representation of the gamma matrices, and $I_{\mathfrak{D}}$ is the $\mathfrak{D} \times \mathfrak{D}$ identity matrix. The Clifford algebra on $M$, denoted by Cl$(M)$, is generated by gamma matrices and their products and contains the chirality matrix 
\begin{align}
\label{eq: Definition of gamma d+1}
    \gamma^{d+1} = i^{\lfloor \frac{d}{2} \rfloor + 1} \gamma^0 \gamma^1 \dots \gamma^{d-1} 
\end{align}
as an element. An explanation for the factor of $i$ and a general discussion of gamma matrices in arbitrary dimension can be found in \cite{SupergravityFreedmanVanproeyen}. In particular, if $d=4$ our chirality matrix is just the usual $\gamma^5$ appearing in the Dirac equation, albeit in a curved spacetime. For a derivation of gamma matrices, frame fields, and spinors in the first order formulation in a curved spacetime, see \cite{CollasKleinDiracGRComputations}. 

Given the fact that $VM$ has dimension $d+1$, it is natural to look for an extension of our gamma matrices to the Thomas cone by adding a single generator. As $M$ is even dimensional we can form such an extension by including $\gamma^{d+1}$ as a generator, following the usual process for dimensionally extending Clifford algebras \cite{SupergravityFreedmanVanproeyen}. Our extended algebra satisfies the commutation relations   
\begin{align}
    \{ \gamma^{d+1} , \gamma^m \} & = 0, \qquad 
    (\gamma^{d+1})^2 = I_{\mathfrak{D}}.
\end{align}
Unfortunately this is not the algebra we are after, as the anticommutators do not give the components of the metric on $VM$. In particular, we are seeking the algebra where 
\begin{align}
    \{ \tilde{\gamma}^M, \tilde{\gamma}^N \} = - 2 G^{MN} I_{\mathfrak{D}}. 
\end{align}
To remedy this situation, we can instead choose $\tilde{\gamma}^N$ to be
\begin{align}
\label{eq: gamma matrices on VM}
    \tilde{\gamma}^n & = \gamma^n, \qquad 
    \tilde{\gamma}^{\lambda} = \tfrac{\lambda}{\lambda_0} ( i \gamma^{d+1} - \lambda_0 g_m \gamma^m ) = \tfrac{\lambda}{\lambda_0} ( \gamma^{\underline{d+1}} - \lambda_0 g_m \gamma^m ), 
\end{align}
(here $\gamma^{\underline{d+1}}$ is the flattened chiral gamma matrix) which satisfy 
\begin{align}
    \{ \tilde{\gamma}^{\lambda}, \tilde{\gamma}^m \} & = 2 \lambda g^{mn} g_n I_{\mathfrak{D}}, \qquad (\tilde{\gamma}^{\lambda})^2 = - \big( \tfrac{\lambda^2}{\lambda_0^2} + \lambda^2 g^{mn} g_m g_n \big) I_{\mathfrak{D}}. 
\end{align}
This set of gamma matrices gives the desired Clifford algebra Cl$(VM)$, wherein we were required to `twist' the last gamma matrix in accordance with the procedure of lifting to $VM$. Of particular importance is understanding the roles of the respective matrices; the matrix $\gamma^{d+1}$ is the usual spacetime chirality matrix. The matrix $\tilde{\gamma}^{\lambda}$ does not have this interpretation as it is defined by the lifting from $M$ to $VM$. The remarkable fact that $\tilde{\gamma}^{\lambda}$ contains $\gamma^{d+1}$ as a component demonstrates that projective geometry provides a natural source of fermion chirality, rather than as an \textit{ad hoc} addition. For an overview of spinors in curved spacetime relevant to the current context, see \cite{YepezReletivisticVierbiens,CollasKleinDiracGRComputations}. For further background of spacetime chirality, see \cite{ AdakNonminimallyCoupledDiracTorsion}. 

Our quest to understand spinors will necessitate the use of indices living in flat spacetime. Using the frame fields we can translate the algebra on $M$ to $TM$ by taking 
\begin{align}
    \gamma^{\underline{a}} = e^{\underline{a}}_{\ m} \gamma^m.
\end{align}
The algebra CL$(\underline{M})$ of flattened gamma matrices satisfies 
\begin{align}
    \{ \gamma^{\underline{a}}, \gamma^{\underline{b}} \} = - 2 \eta^{{\underline{a}} {\underline{b}}} I_{\mathfrak{D}}, 
\end{align}
and contains $\gamma^{\underline{d+1}}$, the usual chiral matrix in flat space. Likewise, taking $\tilde{\gamma}^{\underline{A}} = \tilde{e}^{\underline{A}}_{\ \ M} \tilde{\gamma}^M$ gives the algebra CL$(\underline{VM})$ of flattened gamma matrices on $VM$, obeying 
\begin{align}
    \{ \tilde{\gamma}^{\underline{A}}, \tilde{\gamma}^{\underline{B}} \} = - 2 \eta^{{\underline{A}} {\underline{B}}} I_{\mathfrak{D}}. 
\end{align}
Warning: $\eta^{{\underline{A}} {\underline{B}}}$ is the metric on the tangent bundle of $VM$ and is not the same as the extended metric for $VM$ over a Minkowski spacetime. Explicitly, the extended metric over Minkowski is $\eta^{AB} = \text{diag} (-1,1, \dots, \tfrac{\lambda^2}{\lambda_0^2})$, while the tangent metric is $\eta^{\underline{A} \underline{B}} = \text{diag} (-1,1, \dots, 1)$; only $\eta^{AB}$ has $\lambda$ dependence. 

With all of the gamma matrices in place, it is helpful to organize them by their respective Clifford algebras. The usual algebra CL$(M)$ is generated by the matrices $\gamma^0, \dots, \gamma^{d-1}$ and contains $\gamma^{d+1}$ as an element by virtue of equation \ref{eq: Definition of gamma d+1}, thus having a total dimension of $2^d$. A frequent extension of this algebra is constructed by repurposing $\gamma^{d+1}$ as a generator, giving a Clifford algebra of dimension $2^{d+1}$ \cite{SupergravityFreedmanVanproeyen}. This is not the algebra CL$(VM)$. Rather, the Clifford algebra on $VM$ is generated by products of $\tilde{\gamma}^0, \dots \tilde{\gamma}^{d-1}$ as well as $\tilde{\gamma}^{\lambda}$. This can be identified with the usual extended algebra by exhanging $\tilde{\gamma}^{\lambda} \iff \gamma^{d+1}$. However, this identification really misses the spirit as the algebra on $VM$ is `geometrically twisted' in the sense of equation \ref{eq: gamma matrices on VM}, so we will still distinguish between these choices of generators. 

The flattened Clifford algebra CL$(\underline{M})$ is the usual algebra in Minkowski space found in any quantum field theory text, namely, the algebra generated by the $\gamma^{\underline{a}}$. It contains the flattened chiral gamma matrix $\gamma^{\underline{d+1}}$  as an element. The extension of this latter algebra is obtained by letting $\gamma^{\underline{d+1}}$ be an independent generator. To illuminate and summarize the Clifford algebras and their respective roles, we include the following diagram: 
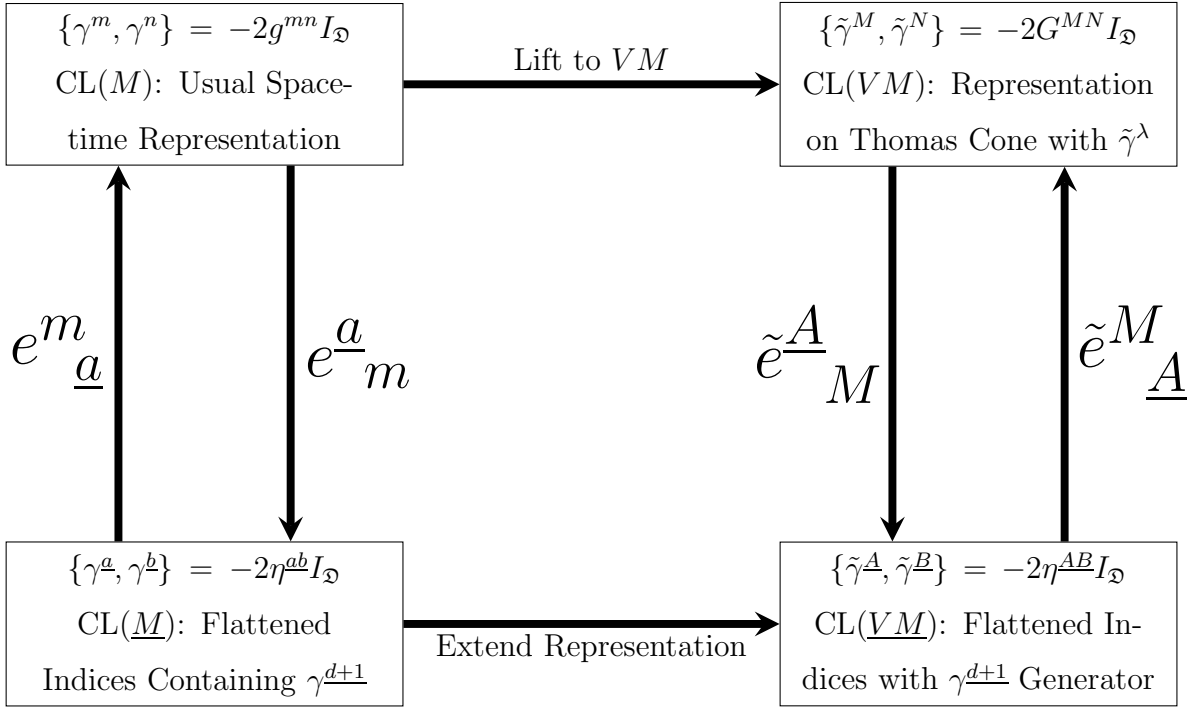
\begin{center}
\begin{tikzpicture}[node distance=5cm, auto]
  \node (top_left) [rectangle, draw, align=center, text width=5cm, minimum height=2cm] {$\{ \gamma^m, \gamma^n \} = - 2 g^{mn} I_\mathfrak{D}$ \\ CL$(M)$:  Usual Spacetime Representation};
  \node (top_right) [rectangle, draw, align=center, text width=5cm, minimum height=2cm, right= of top_left] {$\{ \tilde{\gamma}^M, \tilde{\gamma}^N \} = - 2 G^{MN} I_{\mathfrak{D}}$ \\ CL$(VM)$: Representation on Thomas Cone with $\tilde{\gamma}^{\lambda}$};
  \node (bottom_left) [rectangle, draw, align=center, text width=5cm, minimum height=2cm, below= of top_left] {$\{ \gamma^{\underline{a}}, \gamma^{\underline{b}} \} = - 2 \eta^{{\underline{a}} {\underline{b}}} I_{\mathfrak{D}}$ \\ CL$(\underline{M})$: Flattened Indices Containing $\gamma^{\underline{d+1}}$};
  \node (bottom_right) [rectangle, draw, align=center, text width=5cm, minimum height=2cm, right= of bottom_left] {$\{ \tilde{\gamma}^{\underline{A}}, \tilde{\gamma}^{\underline{B}} \} = - 2 \eta^{{\underline{A}} {\underline{B}}} I_{\mathfrak{D}}$ \\ CL$(\underline{VM})$: Flattened Indices with $\gamma^{\underline{d+1}}$ Generator};

    \draw[>=stealth, ->, line width=3pt] (top_left) -- (top_right) node[shift={(0,0cm)}, pos=.5, align=center] {Lift to $VM$};
    \draw[>=stealth, ->, line width=3pt, font = \huge] (top_left.south east) ++ (-1.5cm, 0cm) -| ([xshift = -1.5cm, yshift = 0cm] bottom_left.north east) node[shift={(0,-2.5cm)}, pos=.5, align=center] {$e_{\ \, m}^{\underline{a}}$};
    \draw[>=stealth, ->, line width=3pt, font = \huge] (bottom_left.north west) ++ (1.5cm, 0cm) -| ([xshift = 1.5cm, yshift = 0cm] top_left.south west) node[shift={(0,2.5cm)}, pos=.5, align=center] {$e^m_{\ \; \underline{a}}$};
    \draw[>=stealth, ->, line width=3pt, font = \huge] (top_right.south west) ++ (1.5cm, 0cm) -| ([xshift = 1.5cm, yshift = 0cm] bottom_right.north west) node[shift={(-2.03cm,-2.5cm)}, pos=.5, align=center] {$\tilde{e}_{\ \ M}^{\underline{A}}$};
    \draw[>=stealth, ->, line width=3pt, font=\huge] (bottom_right.north east) ++ (-1.5cm, 0cm) -| ([xshift = -1.5cm, yshift = 0cm] top_right.south east) node[shift={(1.83cm,2.5cm)}, pos=.5, align=center] {$\tilde{e}^M_{\ \ \underline{A}}$};
    \draw[>=stealth, ->, line width=3pt] (bottom_left) -- (bottom_right) node[shift={(0,-.7cm)}, pos=.5, align=center] {Extend Representation};
\end{tikzpicture}
\end{center}

\subsection{Spin Connection and Spin Representations} 

\label{Section: Spin Connection and Covariant Derivative}

\subsubsection{Spin Connection}

The total covariant derivative of the frame fields (both flat and curved indices) should always vanish \cite{SpinorStuffPoplawski}. Using this fact, we find the spin connection on $VM$  is required to take the form 
\begin{align} \label{eq:Spin Connection}
    \tilde{\omega}_{M \underline{A} \underline{B}} & = \tilde{e}_{\ \ N}^{\underline{C}} \eta_{\underline{A} \underline{C} } \tilde{\nabla}_M \tilde{e}^N_{\ \underline{B}} = \tilde{e}_{\ \ N}^{\underline{C}} \eta_{\underline{A} \underline{C} } \big( \partial_M \tilde{e}^N_{\ \underline{B}} + \tilde{\Gamma}^N_{\ MP} \tilde{e}^P_{\ \underline{B}} \big),  
\end{align}
where $\tilde{\nabla}$ acts only on the unbarred indices. Expanding this expression with the components of the tetrads in equation \ref{eq: frame fields on VM} and $\tilde{\Gamma}$ from equation \ref{eq: TW Connection with torsion}, we find the components of the spin connection are  
\begin{align}
    \tilde{\omega}_{m \underline{a} \underline{b}} & = e_{\ n}^{\underline{c}} \eta_{\underline{a} \underline{c}} \big( \partial_m e^n_{\ \underline{b}} + \Pi^n_{\ mp} e^p_{\ \underline{b}} - (\delta^n_{\ m} - \mathfrak{u}^n_{\ m} ) e^{p}_{\ \underline{b}} g_p \big) \nonumber \\ 
        \tilde{\omega}_{m \underline{a} \underline{\lambda}} & = \tfrac{1}{\lambda_0} e_{\ n}^{\underline{c}} \eta_{\underline{a} \underline{c} } ( \delta^n_{\ m} - \mathfrak{u}^n_{\ m} ) \nonumber \\ 
    \tilde{\omega}_{m \underline{\lambda} \underline{b}} & = \lambda_0 e^{p}_{\ \underline{b}} \big(  g_n \Pi^n_{\ mp} - g_p ( g_m - g_n \mathfrak{u}^n_{\ m} - \mathfrak{a}_m ) - \partial_m g_p + \mathcal{D}_{mp} + \mathcal{V}_{mp} \big) \nonumber \\ 
        \tilde{\omega}_{m \underline{\lambda} \underline{\lambda}} & =  g_m - g_n \mathfrak{u}^n_{\ m} - \mathfrak{a}_m \nonumber \\ 
      \tilde{\omega}_{\lambda \underline{a} \underline{b}} & = \tfrac{1}{\lambda} e_{\ n}^{\underline{c}} \eta_{\underline{a} \underline{c} } ( e^n_{\ \underline{b}} + \mathfrak{u}^n_{\ p} e^p_{\ \underline{b}} ) \nonumber \\ 
        \tilde{\omega}_{\lambda \underline{a} \underline{\lambda}} & = 0 \nonumber \\ 
    \tilde{\omega}_{\lambda \underline{\lambda} \underline{b}} & = \tfrac{\lambda_0}{\lambda} e^p_{\ \underline{b}} ( g_n \mathfrak{u}^n_{\ p} + \mathfrak{a}_p ) \nonumber \\
        \tilde{\omega}_{\lambda \underline{\lambda} \underline{\lambda}} & = \tfrac{1}{\lambda}.
\end{align}
It is worth noting that in our convention $\eta_{\underline{\lambda} \underline{\lambda}} = 1,$ in contrast to \cite{Brensinger:2017gtb,Brensinger:2020gcv,Brensinger:2019mnx} where it is negative, a distinction that carries implications for minus signs in various components of $\tilde{\omega}$. When the torsion fields vanish the components of $\tilde{\omega}$ reduce to 
\begin{align}
    \tilde{\omega}_{m \underline{a} \underline{b}} & = e_{\ n}^{\underline{c}} \eta_{\underline{a} \underline{c}} \big( \partial_m e^n_{\ \underline{b}} + \Pi^n_{\ mp} e^p_{\ \underline{b}} - \delta^n_{\ m} e^{p}_{\ \underline{b}} g_p \big) \nonumber \\ 
        \tilde{\omega}_{m \underline{a} \underline{\lambda}} & = \tfrac{1}{\lambda_0} e_{\ m}^{\underline{c}} \eta_{\underline{a} \underline{c} } \nonumber \\ 
    \tilde{\omega}_{m \underline{\lambda} \underline{b}} & = \lambda_0 e^{p}_{\ \underline{b}} \big(  g_n \Pi^n_{\ mp} - g_p g_m - \partial_m g_p + \mathcal{D}_{mp} \big) \nonumber \\ 
        \tilde{\omega}_{m \underline{\lambda} \underline{\lambda}} & =  g_m \nonumber \\ 
      \tilde{\omega}_{\lambda \underline{a} \underline{b}} & = \tfrac{1}{\lambda} \eta_{\underline{a} \underline{b} } \nonumber \\ 
        \tilde{\omega}_{\lambda \underline{a} \underline{\lambda}} & = \tilde{\omega}_{\lambda \underline{\lambda} \underline{b}} = 0 \nonumber \\ 
        \tilde{\omega}_{\lambda \underline{\lambda} \underline{\lambda}} & = \tfrac{1}{\lambda}.
\end{align}
It provides peace of mind to confirm these are indeed the same coefficients as those derived in \cite{Brensinger:2020gcv}. Further, it is apparent that the torsion fields will induce various fermion couplings not seen in the torsionfree case, as will be elaborated below.

\subsubsection{Thomas Cone Spinors}

An important feature of spinors on $VM$ is their natural relationship with weighted spinors over $M$. Under a Lorentz transformation \(\Lambda^{\underline{a}}_{\ \underline{b}}\), an ordinary spacetime spinor field \(\psi(x)\) on $M$ transforms by
\begin{align} 
\label{eq:ST Spinor}
    \psi \to S \psi, 
\end{align}
where \(S\) is a member of the standard spin representation of the Lorentz group (here we have suppressed spinor indices).

Above we saw the gamma matrices on the Thomas cone are the usual spacetime gamma matrices augmented by \(\tilde{\gamma}^{\lambda}\), implying spinors on $VM$ have the same dimension as those on $M$. Therefore spinors on $VM$ take the form
\begin{align} 
\label{eq:TW Spinor}
    \Psi(x,\lambda) = \mathcal{W}(x,\lambda)\psi(x), 
\end{align}
where \( \psi = \psi(x)\) is a spinor on \(M\) and \(\mathcal{W}(x,\lambda)\) is an object that acts linearly on \(\psi\) at each point \((x^a, \lambda)\). Furthermore, at any particular fixed value of $\lambda$, say \(\hat{\lambda}\), the field \(\Psi(x,\hat{\lambda})\) must project down to a spinor field on \(M\). This requires \(\mathcal{W}(x,\hat{\lambda})\) to be an element of the spin representation (i.e. a change-of-basis in spinor space) for any \(\hat{\lambda}\), so \(\mathcal{W}(x,\hat{\lambda}) \to S\mathcal{W}(x,\hat{\lambda})S^{-1}\) under Lorentz transformations on M. This leads to the conclusion \(\mathcal{W}(x,\lambda)\) has the form
\begin{align} 
\label{eq:TW Spinor 2}
    \mathcal{W} (x,\lambda) = f(\lambda) W(x), 
\end{align}
where \(W = W(x)\) is an element of the spin representation of the Lorentz group over \(M\). 

We would like to understand what form $f(\lambda)$ can take. The answer is unveiled through the transformation law of \(\Psi(x,\lambda)\). If \(f(\lambda) \to \hat{f}(\lambda)\) under a transformation \(x^a \to y^a, \lambda \to \lambda |J|^{-\frac{1}{d+1}} \), then
\begin{align} 
\begin{split} 
\label{eq:TW Spinor Coordinate Transformation}
    \Psi & = W f(\lambda) \psi \to W \hat{f}(\lambda) \psi = \frac{\hat{f}(\lambda)}{f(\lambda)} \Psi.
    \end{split}
\end{align} 
This transformation must be autonomous in both \(x^a\) and \(\lambda\) (that is, there should be no explicit dependence on $\lambda$), which requires the ratio \(\frac{\hat{f}(\lambda)}{f(\lambda)}\) to have no $\lambda$ dependence. This can be achieved by \(f(\lambda) = \left( \frac{\lambda}{\lambda_0} \right)^{\weight} \) where $\weight$ can be any real number (here \inve \ is the `schwa' character, pronounced `schwa'). With this in place, we finally see 
\begin{align} 
\label{eq:TW Spinor 3}
    \mathcal{W}(x, \lambda) = \left( \tfrac{\lambda}{\lambda_0} \right)^{\weight} W,
\end{align}
which means 
\begin{align} 
\label{eq:TW Spinor 4}
    \Psi(x, \lambda) = \left( \tfrac{\lambda}{\lambda_0} \right)^{\weight} W \psi(x).
\end{align}
Under Lorentz transformations on $M$, the TW spinor of equation \ref{eq:TW Spinor 4} transforms like an ordinary spinor, \(\Psi \to S \Psi\). Further, under a coordinate transformation \(x^a \to y^a\) on $M$, the transformation law for $\Psi = \Psi(x, \lambda)$ on $VM$ is identical to that of a weighted spinor on $M$ with weight $- \frac{\weight}{d+1}$, 
\begin{align} 
\label{eq:TW Spinor Coordinate Transformation 2}
    \Psi \to |J|^{- \frac{\weight}{d+1}} \Psi. 
\end{align}
This statement fixes our convention for weights throughout. We note that our selection for $f (\lambda)$ provides a choice of local sections of $VM$, which raises the possibility of dynamical weights. Combining a Lorentz transformation \(\Lambda^{\underline{A}}_{\ \underline{B}}\) with a TW coordinate transformation gives the total transformation law
\begin{align} 
\label{eq:TW Spinor Coordinate Transformation 3}
    \Psi \to |J|^{- \frac{\weight}{d+1}} S \Psi. 
\end{align}

\subsubsection{Adjoint Spinors}

The adjoint spinor \(\bar{\Psi}(x,\lambda)\) of \(\Psi(x,\lambda)\) can be determined by demanding \(\bar{\Psi}\Psi\) to transform as a scalar under Lorentz transformations and coordinate transformations of $VM$; in other words, \(\bar{\Psi}\Psi \to \bar{\Psi}\Psi\). This necessitates  
\begin{align} 
\label{eq:Adjoint TW Spinor Lorentz Transformation}
    \bar{\Psi} \to \bar{\Psi} S^{-1}
\end{align}
under Lorentz transformations, and
\begin{align} 
\label{eq:Adjoint TW Spinor Coordinate Transformation}
    \bar{\Psi} \to \bar{\Psi} |J|^{\frac{\weight}{d+1}}
\end{align}
under TW coordinate transformations.

As always, the Dirac adjoint \(\bar{\psi}(x)\) of the manifold spinor \(\psi(x)\) transforms as \(\bar{\psi} \to \bar{\psi} S^{-1}\). As with vector fields on $VM$, the inner product should be preserved when lifting from $M$ to $VM$, i.e., we should have $\bar{\psi} \psi = \bar{\Psi} \Psi$. Using the form of $\Psi$ found in equation \ref{eq:TW Spinor 4}, this requires
\begin{align}
    \bar{\Psi} \Psi & = \bar{\Psi} \left( \frac{\lambda}{\lambda_0} \right)^{\weight} W \psi(x) = \bar{\psi} \psi,
\end{align}
which, combined with the transformation laws in equations \ref{eq:Adjoint TW Spinor Lorentz Transformation} and \ref{eq:Adjoint TW Spinor Coordinate Transformation}, fixes $\bar{\Psi}$ to be 
\begin{align} 
\label{eq:Adjoint TW Spinor}
    \bar{\Psi} = \bar{\psi}(x) W^{-1} \left( \frac{\lambda}{\lambda_0} \right)^{- \weight}.
\end{align}
It is straightforward to verify this has the desired transformation laws.

\subsubsection{Representations and Clifford Algebras}

It is worthwhile to transcribe several notes about the coordinate transformation of TW spinors given by equation \ref{eq:TW Spinor Coordinate Transformation 2}. Elements of the spin representation on \(M\), such as \(S\) and \(W\), may be written in terms of generators of the spacetime Clifford algebra via the exponential map. For the present we label the generators as \(v^{\overline{i}}\) (so $\overline{i}$ ranges from $1$ to $2^d$) and write
\begin{align} 
\label{eq:Spin Representative In Terms of Clifford}
    W &= \exp(w_{\overline{i}} v^{\overline{i}}).
\end{align}
Here the weights \(w_{\overline{i}}\) are complex functions on $M$. 

Substituting the components from equation \ref{eq:Spin Representative In Terms of Clifford} in the definition of $\Psi$, we have 
\begin{align} 
\label{eq:TW Spinor In Terms of Clifford}
    \Psi = \left( \tfrac{\lambda}{\lambda_0} \right)^{\weight} \exp(w_{\overline{i}} v^{\overline{i}}) \psi(x).
\end{align}
Moving the $\lambda$ term inside the exponential, we can express spinors on $VM$ in terms of those on $M$ at the level of the algebra. Indeed, 
\begin{align} 
\label{eq:TW Spinor In Terms of Clifford Algebra }
    \Psi = \exp(w_{\overline{i}} v^{\overline{i}} + \weight \log ( \tfrac{\lambda}{\lambda_0} ) I_{\mathfrak{D}} ) \psi(x).
\end{align}
It is also straightforward to understand the transformation law from the algebra. In analogy with the decomposition of $W$ in equation \ref{eq:Spin Representative In Terms of Clifford}, we write $S = \exp (s_{\overline{i}} v^{\overline{i}})$, so the combined transformation law in equation \ref{eq:TW Spinor Coordinate Transformation 3} becomes 
\begin{align} 
\label{eq:TW Spinor Transformation in Terms of Clifford}
    \Psi & \to \exp \left( - \tfrac{\weight}{d+1} \log |J| I_{\mathfrak{D}} + s_{\overline{i}} v^{\overline{i}} \right) \Psi \nonumber \\  
        & = \exp \left( \weight \left( \log \left( \tfrac{\lambda}{\lambda_0} \right) + \log |J|^{- \frac{1}{d+1}} \right) I_{\mathfrak{D}}+ \left( s_{\overline{i}} + w_{\overline{i}} \right) v^{\overline{i}} \right) \psi.
\end{align}
Putting this all together, equation \ref{eq:TW Spinor Transformation in Terms of Clifford} classifies the spinors we may expect to see; spinors on $VM$ have the same transformation as spinors on $M$ with an additional overall weight. Ultimately the dependence on $\tfrac{\lambda}{\lambda_0}$ introduces the weight, whereas the `coupling' is determined by $\weight$.

\subsection{Spinor Covariant Derivative}

Armed with a basic understanding of the properties of TW spinors, we may now investigate their dynamics. From equation \ref{eq:TW Spinor Coordinate Transformation 3}, we see the derivative of \(\Psi\) transforms under a simultaneous Lorentz transformation \(\Lambda^{\underline{A}}_{\ \underline{B}}\) and TW coordinate transformation \(x^a \to y^a\), \(\lambda \to \lambda |J|^{-\frac{1}{d+1}}\) as
\begin{align} 
\begin{split} 
\label{eq:Transformation of TW Spinor Derivative} 
    \partial_A \Psi & \to \frac{\partial x^B}{\partial y^A} \partial_B \left( |J|^{- \frac{\weight}{d+1}} S \Psi \right) = \frac{\partial x^B}{\partial y^A} \left( |J|^{- \frac{\weight}{d+1}} S \partial_B \Psi + \partial_B \left( |J|^{- \frac{\weight}{d+1}} S \right) \Psi \right). \end{split}
\end{align}
Equation \ref{eq:Transformation of TW Spinor Derivative} demonstrates that the partial derivative on $VM$ does not preserve spinorality, in accordance with ordinary spinors \cite{SpinorStuffPoplawski}. As with any geometric object, we can only make meaningful statements about the dynamics of a TW spinor field with a covariant derivative, which we define by 
\begin{align}
    \tilde{\nabla}_M = \partial_M + \tilde{\Omega}_M. 
\end{align}
This serves as the definition of the TW spinor connection \(\tilde{\Omega}_M\). Under Lorentz and TW coordinate transformations, the covariant derivative of a TW spinor transforms as
\begin{align} 
\begin{split} 
\label{eq:Transformation of TW Spinor Covariant Derivative}
    \tilde{\nabla}_A \Psi & = \partial_A \Psi + \tilde{\Omega}_A \to \frac{\partial x^B}{\partial y^A} \left( |J|^{- \frac{\weight}{d+1}} S \partial_B \Psi + \partial_B \left( |J|^{- \frac{\weight}{d+1}} S \right) \Psi \right) + \hat{\tilde{\Omega}}_A |J|^{- \frac{\weight}{d+1}} S \Psi. 
\end{split}
\end{align}
We will determine the form of the transformed coefficients \(\hat{\tilde{\Omega}}_A\) out of a desire for \(\tilde{\nabla}_A\Psi\) to have the same transformation law as \(\Psi\) with an additional vector index. In other words, it should be the case that
\begin{align} 
\label{eq:Transformation of TW Spinor Covariant Derivative 2}
    \tilde{\nabla}_A \Psi \to \frac{\partial x^B}{\partial y^A} |J|^{- \frac{\weight}{d+1}} S \tilde{\nabla}_B \Psi.
\end{align}
Comparing equations \ref{eq:Transformation of TW Spinor Covariant Derivative} and \ref{eq:Transformation of TW Spinor Covariant Derivative 2}, we find that
\begin{align} 
\label{eq:Transformation of TW Spinor Connection}
    \hat{\tilde{\Omega}}_A & = \frac{\partial x^B}{\partial y^A} \left( S \tilde{\Omega}_B S^{-1} - \partial_B \left( |J|^{-\frac{\weight}{d+1}} S \right) S^{-1} |J|^{\frac{\weight}{d+1}} \right) \nonumber \\ 
    & = \frac{\partial x^B}{\partial y^A} \left( S \tilde{\Omega}_B S^{-1} - \left( \partial_B S \right) S^{-1} - \partial_B \left( |J|^{- \frac{\weight}{d+1}} \right) |J|^{\frac{\weight}{d+1}} \right). 
\end{align}
This is a desirable transformation law to obtain; indeed, examining equation \ref{eq:Transformation of TW Spinor Connection}, we see $\tilde{\Omega}_A$ has the same transformation as a connection associated with a spinor density of weight \( - \frac{\weight}{d+1}\) on \(M\). Breaking equation \ref{eq:Transformation of TW Spinor Connection} into its spacetime and \(\lambda\) components, we have the transformation laws
\begin{align}
\label{eq:Transformation of TW Spinor Connection in Components}
    \tilde{\Omega}_a & \to \bar{J}^b_{\ a} \left( S \left( \tilde{\Omega}_b - \lambda j_b \tilde{\Omega}_{\lambda} \right) S^{-1} - \left( \partial_b S  \right) S^{-1} -\weight j_b \right) \nonumber \\ 
    \tilde{\Omega}_{\lambda} & \to S \left( |J|^{\frac{1}{d+1}} \tilde{\Omega}_{\lambda} \right) S^{-1}. 
\end{align}

\subsection{The Spinor Connection in Terms of the Spin Connection}

It has long been known that the presence of torsion in a gravitational theory can significantly alter the behavior of spinors \cite{SpinAndTorsion,DiracEquationTorsionZecca}, and thus the spinor covariant derivative \cite{DiracEquationTorsionNonMetricityADR}. At this point it is reasonable to hypothesize that there may be a similar effect in TW gravity, with the additional possibility of the TW fields influencing fermonic physics. To pursue this investigation it would be advantageous to obtain an explicit description of the TW spinor connection $\tilde{\Omega}_A$, which can be achieved by relating $\tilde{\Omega}_A$ to the TW spin connection  $\tilde{\omega}_{C \underline{A} \underline{B}}$. This derivation follows from the requirement that the total covariant derivative (i.e. respecting both curved and flat indices) of the frame fields is zero, a condition equivalent to equation \ref{eq:Spin Connection}. Since \(\bar{\Psi}\Psi\) is a scalar under TW coordinate transformations and TW Lorentz transformations, we have
\begin{equation}
    \tilde{\nabla}_A \left( \bar{\Psi}\Psi \right) = \partial_A \left( \bar{\Psi}\Psi \right).
\end{equation}
Expanding both sides of this equation with the Leibniz rule and simplifying, we arrive at the expression
\begin{equation}
    \tilde{\nabla}_A \bar{\Psi} = \partial_A \bar{\Psi} - \bar{\Psi} \tilde{\Omega}_A,
\end{equation}
where $\tilde{\Omega}_A$ is applied on the right and with a minus sign. 

Next we examine the covariant derivative of the metric tensor \(G_{AB}\) and gamma matrices \(\tilde{\gamma}^A\). We denote the non-metricity tensor of the TW connection by \(\tilde{N}_{CAB}\), 
\begin{equation}
    \tilde{\nabla}_C G_{AB} = \tilde{N}_{CAB}.
\end{equation}
The defining relation of the gamma matrices on $VM$ is \(\tilde{\gamma}^A\tilde{\gamma}^B + \tilde{\gamma}^B\tilde{\gamma}^A = -2G^{AB}\). Taking the covariant derivative of both sides and simplifying leads to
\begin{equation}
    \tilde{\nabla}_C \tilde{\gamma}^A = \tfrac{1}{2} \tilde{N}_{ C \ B}^{\ A} \tilde{\gamma}^B.
\end{equation}

We now have all of the necessary pieces to find an expression for \(\tilde{\Omega}_B\). The object $\bar{\Psi} \tilde{\gamma}^A \Psi$ transforms as a vector under coordinate transformations \(x^a \to y^a, \lambda \to \lambda |J|^{-\frac{1}{d+1}} \) on $VM$, meaning
\begin{equation}  \label{eq:Partial and Covariant Derivatives of Dirac Vector 1}
    \tilde{\nabla}_B \left( \bar{\Psi} \tilde{\gamma}^A\Psi \right) = \partial_B \left( \bar{\Psi} \tilde{\gamma}^A \Psi \right) + \tilde{\Gamma}^A_{\ BC} \left( \bar{\Psi} \tilde{\gamma}^C \Psi \right).
\end{equation}
Expanding this equation with the Leibniz rule and collecting factors of $\Psi$ gives
\begin{align} 
\begin{split}
\label{eq:Partial and Covariant Derivatives of Dirac Vector 1}
    & \bar{\Psi} \left( - \tilde{\Omega}_B \tilde{\gamma}^A + \tfrac{1}{2} \tilde{N}_{C \ B}^{\ A} \tilde{\gamma}^C + \tilde{\gamma}^A \tilde{\Omega}_B \right) \Psi = \bar{\Psi} \tilde{\omega}_{B \ \underline{C}}^{\ A} \tilde{\gamma}^{\underline{C}} \Psi, 
\end{split} 
\end{align}
from which it follows that 
\begin{align} 
\begin{split} 
\label{eq:Partial and Covariant Derivatives of Dirac Vector 1.5}
    & \tilde{\gamma}^{\underline{E}} \tilde{\Omega}_B - \tilde{\Omega}_B \tilde{\gamma}^{\underline{E}} + \tfrac{1}{2} \tilde{N}_{B \ \underline{C}}^{\ \underline{E}} \tilde{\gamma}^{\underline{C}} = \tilde{\omega}_{B \ \underline{C}}^{\ \underline{E}}\tilde{\gamma}^{\underline{C}}.
\end{split} 
\end{align}
Rearranging a few terms and contracting both sides with $\tilde{\gamma}_E$, we conclude 
\begin{align}
    \label{eq: CommutatorOmegaDefinition} \tilde{\gamma}_{\underline{E}} \left[ \tilde{\Omega}_B, \tilde{\gamma}^{\underline{E}} \right] = \tfrac{1}{2} \tilde{N}^{\underline{E}}_{\ \underline{E} B} - \tilde{\omega}_{B \underline{E} \underline{C}} \tilde{\gamma}^{\underline{E}} \tilde{\gamma}^{\underline{C}}.
\end{align}
We now take the ansatz
\begin{equation}
\label{eq:General Form of Big Omega in Terms of little omega}
    \tilde{\Omega}_B = \tfrac{1}{d+1} \tilde{\omega}_{B[\underline{E}\underline{C}]}\tilde{\gamma}^{\underline{E}}\tilde{\gamma}^{\underline{C}} + V_B 
\end{equation}
where \(V_B\) is some spinor object with one (not necessarily tensorial) $VM$ index. To solve for \(V_B\), we plug the ansatz from equation \ref{eq:General Form of Big Omega in Terms of little omega} back into equation \ref{eq: CommutatorOmegaDefinition}. After some simplification, this yields
\begin{equation}
    \tilde{\gamma}_{\underline{E}}V_B\tilde{\gamma}^{\underline{E}} - (d+1)V_B = 0,
\end{equation}
from which we conclude that \(V_B\) is in the center of the algebra, and therefore each fixed value $B$ yields a multiple of the spinor identity. This can be seen by noting \(\tilde{\gamma}_{\underline{E}} \tilde{\gamma}^{\underline{E}} = (d+1) I_{\mathfrak{D}} \), which imposes the condition $[V_B, \tilde{\gamma}^{\underline{E}}] = 0$. With this form of \(V_B\), the TW spinor connection coefficients \(\tilde{\Omega}_B\) are given in terms of $\tilde{\omega}_{C\underline{A}\underline{B}}$ by equation \ref{eq:General Form of Big Omega in Terms of little omega}.

Now we restrict the form of \(V_B\) in view of the transformation law of \(\tilde{\Omega}_B\) given by equation \ref{eq:Transformation of TW Spinor Connection}. Letting \(T\) be any scalar density of weight \(-\frac{1}{d+1}\) on \(M\), under TW coordinate transformations we have
\begin{equation}
    \partial_B \log(T^\weight) \to \bar{J}^C_{\ B} \partial_C \log(T^\weight |J|^{-\frac{\weight}{d+1}}) = \bar{J}^C_{\ B}\left( \partial_C \log(T^\weight) + \partial_C \left( |J|^{-\frac{\weight}{d+1}}\right) |J|^{\frac{\weight}{d+1}} \right).
\end{equation}
This transformation is clearly non-tensorial on $VM$ due to the latter term. However, up to a minus sign this is precisely what is needed for the general transformation law of \(\tilde{\Omega}_B\) given by equation \ref{eq:Transformation of TW Spinor Connection}. Therefore, we conclude 
\begin{align}
    V_B = - \partial_B \log \left( T^\weight \right),
\end{align}
and rewrite equation \ref{eq:General Form of Big Omega in Terms of little omega} as
\begin{equation} 
\label{eq:Spinor Connection with T}
    \tilde{\Omega}_B = \tfrac{1}{d+1} \tilde{\omega}_{B[\underline{E}\underline{C}]}\tilde{\gamma}^{\underline{E}}\tilde{\gamma}^{\underline{C}} - \partial_B \log(T^\weight).
\end{equation}
Noting that \(\sqrt{|g|}\) is a scalar density of weight 1, we see that \(\sqrt{|g|}^{-\frac{1}{d+1}}\) is a possible choice for $T$ in equation \ref{eq:Spinor Connection with T} (a scalar density of weight \(- \tfrac{1}{d+1}\)). Plugging this choice of $T$ into equation \ref{eq:Spinor Connection with T} yields
\begin{align} 
\begin{split} 
\label{eq:Spinor Connection with g_b}
    \tilde{\Omega}_b & = \tfrac{1}{d+1} \tilde{\omega}_{b[\underline{E}\underline{C}]}\tilde{\gamma}^{\underline{E}} \tilde{\gamma}^{\underline{C}} - \weight g_b, \\
    \tilde{\Omega}_{\lambda} & = \tfrac{1}{d+1} \tilde{\omega}_{\lambda[\underline{E}\underline{C}]}\tilde{\gamma}^{\underline{E}}\tilde{\gamma}^{\underline{C}}. 
\end{split} 
\end{align}
More generally, \(\tilde{\Omega}_b\) may be offset from the expression given in equation \(\ref{eq:Spinor Connection with g_b}\) by a vector on $M$ times the spinor identity matrix. This amounts to replacing \(g_b\) with \(\kappa_b\). This is the most general form for the TW spinor connection \(\tilde{\Omega}_B\) in terms of the TW spin connection $\tilde{\omega}_{C\underline{A}\underline{B}}$. In this paper, we focus on the specific case where \(g_b\) appears in the spinor connection, which amounts to assuming the Roberts gauge.

Finally, note what happens if we include a term with the symmetric part of the spin connection coefficients; i.e., \(\frac{1}{d+1}\tilde{\omega}_{C(\underline{A}\underline{B})}\tilde{\gamma}^{\underline{A}}\gamma^{\underline{B}}\). Since \(\tilde{\omega}_{C(\underline{A}\underline{B})}\) is proportional to the non-metricity tensor \cite{SpinorStuffPoplawski,GaugeTheoryGravitySpacetimeHehl}, this term is a vector multiple of the identity matrix in the spinor space. In light of the previous discussion, this means that this term will simply have the same effect as shifting \(g_b\) to some \(\kappa_b\) in equation \ref{eq:Spinor Connection with g_b}. Therefore such a term can be included in the spinor connection \(\tilde{\Omega}_{B}\), but it is not as fundamental as the antisymmetric part, which is necessary to produce the correct transformation law.

\subsection{TW Dirac Equation}

Now that all of the geometric preliminaties are established we are ready to present the TW Dirac equation. Based on knowledge of spinors in general relativity \cite{DiracOperators}, we expect to find torsion contributing to spin interactions with fermions \cite{DiracEquationTorsionZecca, InteractingGravityAndDiracZecca, SpinAndTorsion}. Defining the differential operator 
\begin{align}
\label{eq: unexpanded baby Dirac operator}
    \tilde{\nabla}_M & = \partial_M + \tilde{\Omega}_M, \nonumber \\ 
    \tilde{\slashed{\nabla}} & = \tilde{\gamma}^M \tilde{\nabla}_M, 
\end{align}
the Dirac equation on $VM$ is 
\begin{align}
\label{eq: unexpanded baby Dirac equation}
    \left( i \tilde{\slashed{\nabla}} - m \right) \Psi = 0. 
\end{align}
We acknowledge that to preserve unitarity the metric-affine Dirac equation should have the term $\tilde{\nabla}_M (\tilde{\gamma}^M \Psi)$ instead of $\tilde{\gamma}^M \tilde{\nabla}_M \Psi$ \cite{GaugeTheoryGravitySpacetimeHehl}. However, the part we have not included is 
\begin{align}
\label{eq: derivative of gamma in Dirac equation}
    \left( \tilde{\nabla}_M \gamma^M \right) \Psi = \tfrac{1}{2} \tilde{N}_{ M \ B}^{\ M} \tilde{\gamma}^B \Psi,
\end{align}
which can be recovered by the replacement \(\tilde{\Omega}_M \to \tilde{\Omega}_M + \tfrac{1}{2} \tilde{N}_{ A \ M}^{\ A}\). This amounts to adding a vector multiple of the spinor space identity matrix to the spinor connection, or equivalently, replacing $g_b$ with an arbitrary $\kappa_b$ in equation \ref{eq:Spinor Connection with g_b}. Since we are already restricting to the Roberts gauge $\kappa_b=g_b$ in this paper, there is no reason to include this term in the Dirac equation.

Expanding $\tilde{\slashed{\nabla}}$ by substituting the gamma matrices from equation \ref{eq: gamma matrices on VM} and the components of $\tilde{\Omega}$ from equation \ref{eq:Spinor Connection with g_b}, the spinorial influence of $\tilde{\slashed{\nabla}}$ is revealed. Indeed, equation \ref{eq: unexpanded baby Dirac equation} can equivalently be written 
\begin{align}
\label{eq: Dirac equation on VM}
    \left( i \slashed{\partial} + \tfrac{i}{2(d+1)} \left( \left( \slashed{\partial} e^n_{\ \underline{b}} \right) \gamma_{n}^{\  \underline{b}} + \mathcal{O} - 2 i \left( \chi + \tfrac{d - \mathfrak{u}}{\lambda_0} \right) \gamma^{\underline{d+1}} \right) + i \tilde{\gamma}^{\lambda} \Xi_{\lambda} - m \right) \Psi = 0, 
\end{align}
where $\slashed{\partial} = \gamma^m \partial_m$ denotes the usual Dirac operator on $M$, and we have introduced $\mathcal{O}, \chi$, and $\Xi_{\lambda}$ as 
\begin{align}
    \mathcal{O} & = \gamma^m \left( \Pi_{nmp} - g_p \left( g_{nm} - \mathfrak{u}_{nm} \right) \right) \left( \gamma^n \gamma^p - \gamma^p \gamma^n \right) - \weight \gamma^m g_m,  \nonumber \\ 
    \chi & = \lambda_0 \left( g_n \Pi^n_{\ mp} - g_p \left( g_m - g_n \mathfrak{u}^n_{\ m} - \mathfrak{a}_m \right) +  \partial_m g_p - \mathcal{D}_{mp} + \mathcal{V}_{mp} \right) \gamma^m \gamma^p, \quad \text{and} \nonumber \\ 
    \Xi_{\lambda} & = \partial_{\lambda} - \tfrac{\lambda_0}{\lambda (d+1)} \left( g_n \mathfrak{u}^n_{\ p} + \mathfrak{a}_p \right) \gamma^p \gamma^{\underline{d+1}}.
\end{align}
For a free spinor field the first term $\mathcal{O}$ is an ordinary mass term, while the factor of $\chi$ contributes to a $\chi$ral (chiral) mass. The presence of the TW fields in $\chi$ also indicates that the TW fields contribute to a chiral mass. At this stage the role of $\Xi$ is not so clear, but we will shortly find a better approach to understanding its physical contribution. Taking the torsionfree limit, the above reduces to 
\begin{align}
\label{eq: Dirac equation on VM without torsion}
    \left( i \slashed{\partial} + \tfrac{i}{2(d+1)} \left( \left( \slashed{\partial} e^n_{\ \underline{b}} \right) \gamma_n^{\  \underline{b}} + \mathcal{O} - 2 i \left( \chi + \tfrac{d}{\lambda_0} \right) \gamma^{\underline{d+1}} \right) + i \tilde{\gamma}^{\lambda} \partial_{\lambda} - m \right) \Psi = 0, 
\end{align}
where now 
\begin{align}
    \mathcal{O} & = \gamma^m \left( \Pi_{nmp} - g_p g_{nm} \right) \left( \gamma^n \gamma^p - \gamma^p \gamma^n \right) - \weight \gamma^m g_m  \quad \text{and} \nonumber \\ 
    \chi & = \lambda_0 \left( g_n \Pi^n_{\ mp} - g_p g_m +  \partial_m g_p - \mathcal{D}_{mp} \right) g^{mp}. 
\end{align}
Interestingly, the factor of $\Xi$ has now reduced to a $\lambda$ derivative operator. 

Although equations \ref{eq: Dirac equation on VM} and \ref{eq: Dirac equation on VM without torsion} provide a concise theoretical formulation for studying the Dirac equation on $VM$, their experimental interpretations leave much to be desired, as any observable physical quantity lives on $M$ \cite{Brensinger:2020gcv}. We therefore want to expand $\Psi$ into manifold components, under which equation \ref{eq: Dirac equation on VM} becomes 
\begin{align}
    \left( i \slashed{\partial} + \tfrac{i}{2(d+1)} \left( \left( \slashed{\partial} e^n_{\ \underline{b}} \right) \gamma_n^{\  \underline{b}} + \mathcal{O} - 2 i \left( \chi + \tfrac{d - \mathfrak{u}}{\lambda_0} \right) \gamma^{\underline{d+1}} \right) + \tfrac{i}{\lambda} \tilde{\gamma}^{\lambda} \Xi - m \right) \left( W \psi \right) = 0, 
\end{align}
where now 
\begin{align}
    \mathcal{O} & = \gamma^m \left( \Pi_{nmp} - g_p \left( g_{nm} - \mathfrak{u}_{nm} \right) \right) \left( \gamma^n \gamma^p - \gamma^p \gamma^n \right) - \weight \gamma^m g_m,  \nonumber \\ 
    \chi & = \lambda_0 \left( g_n \Pi^n_{\ mp} - g_p \left( g_m - g_n \mathfrak{u}^n_{\ m} - \mathfrak{a}_m \right) +  \partial_m g_p - \mathcal{D}_{mp} + \mathcal{V}_{mp} \right) \gamma^m \gamma^p, \quad \text{and} \nonumber \\ 
    \Xi & = \weight - \tfrac{\lambda_0}{(d+1)} \left( g_n \mathfrak{u}^n_{\ p} + \mathfrak{a}_p \right) \gamma^p \gamma^{\underline{d+1}}.
\end{align}
The mass $\mathcal{O}$ and chiral mass $\chi$ contributions are unchanged from before, but $\Xi$ no longer contains a $\lambda$ derivative operator, and is now a scalar. In fact, the $\partial_{\lambda}$ term has sourced the explicit $\weight$ factor now appearing in $\Xi$. Again taking the torsionfree limit gives 
\begin{align}
    \left( i \slashed{\partial} + \tfrac{i}{2(d+1)} \left( \left( \slashed{\partial} e^n_{\ \underline{b}} \right) \gamma_n^{\  \underline{b}} + \mathcal{O} - 2 i \left( \chi + \tfrac{d}{\lambda_0} \right) \gamma^{\underline{d+1}} \right) + \tfrac{i \weight}{\lambda} \tilde{\gamma}^{\lambda} - m \right) \left( W \psi \right) = 0, 
\end{align}
with 
\begin{align}
    \mathcal{O} & = \gamma^m \left( \Pi_{nmp} - g_p g_{nm} \right) \left( \gamma^n \gamma^p - \gamma^p \gamma^n \right) - \weight \gamma^m g_m \quad \text{and}  \nonumber \\ 
    \chi & = \lambda_0 \left( g_n \Pi^n_{\ mp} - g_p g_m +  \partial_m g_p - \mathcal{D}_{mp} \right) \gamma^m \gamma^p. 
\end{align}
This framework for studying the Dirac equation is more general than has been previously considered. However, the above equations are still restricted to the Roberts gauge. The dynamics of the Dirac equation when allowing this full gauge symmetry is under investigation.

\section{Maxwell Interactions}

\subsection{Thomas-Whitehead Action}

Before undertaking a study of gauge fields in TW gravity we summarize the TW action and several relevant features. The Thomas-Whitehead action \cite{Brensinger:2017gtb, Brensinger:2019mnx, Brensinger:2020gcv} consists of two components, the Projective Einstein-Hilbert action $S_{\text{PEH}}$ and the Projective Gauss-Bonnet action $S_{\text{PGB}}$, which are given by 
\begin{equation} 
    \label{eq:PEH Action} 
    S_{\text{PEH}} = \frac{1}{2 \kappa_0} \int \sqrt{|G|} \mathcal{K} d \lambda d^d x,  
\end{equation}
and 
\begin{equation} 
    \label{eq:PGB Action} S_{\text{PGB}} = J_0 c \int \sqrt{|G|} \left(\mathcal{K}^{A}_{\ BCD} \mathcal{K}_{A}^{\ BCD} - 4 \mathcal{K}_{AB} \mathcal{K}^{AB} + \mathcal{K}^2 \right) d \lambda d^d x, 
\end{equation}
where the whole TW action is then $S_{\text{TW}} = S_{\text{PEH}} + S_{\text{PGB}}$. In the TW action $S_{\text{PEH}}$ couples the metric to $\mathcal{D}$ and $\overline{C}$, whereas $S_{\text{PGB}}$ endows $\mathcal{D}$ and $\overline{C}$ with dynamics through the Riemann squared and Ricci squared terms. The fact that the TW action takes the form of $S_{\text{PEH}} + S_{\text{PGB}}$ avoids higher derivatives that could lead to instabilities \cite{Lovelock:1971yv, Lanczos:1938sf}. From equations \ref{eq:PEH Action} and \ref{eq:PGB Action}, we can see that the only components present in the TW action are built from $\mathcal{K}^A_{\ BCD}$, $\mathcal{K}_{BD}$, and $\mathcal{K}$. Henceforth, in our study of electromagnetic and Yang-Mills couplings, we will be interested not in the exact form of the action so much as the contractions of $\mathcal{K}^A_{\ BCD}$ that appear. 

As mentioned above, the Projective Gauss-Bonnet term renders $\mathcal{D}$ and $\overline{C}$ as dynamical fields. Conceivably there are many Lagrangians that could be paired with $S_{\text{PEH}}$ that would accomplish this purpose. The motivation behind the particular choice of $S_{\text{PGB}}$ lies in the property of weak field approximations. We define the $\textit{Lovelock limit}$ of $S_{\text{TW}}$ (after \cite{Lovelock:1971yv}) to be the result of taking both $\mathcal{D}, \overline{C} \to 0$, which may equivalently be though of as a vanishing diff field and choosing $\Gamma^a_{\ bc}$ to be the usual Levi-Civita connection $\hat{\Gamma}^a_{\ bc}$. We then further define the $\textit{Einstein limit}$ to be the Lovelock limit with $J_0 \to 0$. In the Lovelock limit the action becomes (up to an overall integral of  $\tfrac{1}{\lambda}$)  
\begin{equation} 
    \label{eq:Lovelock limit} 
    S_{\text{Lovelock}} = \frac{1}{2 \kappa_0} \int \sqrt{|g|} \hat{R} \, d^d x + J_0 c \int \sqrt{|g|} \ \left( \hat{R}^a_{\ bcd} \hat{R}_a^{\ bcd} - 4 \hat{R}_{ab} \hat{R}^{ab} + \hat{R}^2 \right) d^d x. 
\end{equation}
The first term is just the usual Einstein-Hilbert action, while the second is the Gauss-Bonnet action, which by virtue of the Chern-Gauss-Bonnet theorem does not contribute to the field equations in $d \leq 4$. Further applying the Einstein limit ensures this term does not contribute in any dimension and recovers the action 
\begin{equation} 
    \label{eq:Einstein limit} 
    S_{\text{Einstein-Hilbert}} = \frac{1}{2 \kappa_0} \int \sqrt{|g|} \hat{R} \, d^d x. 
\end{equation}
In other words, the Einstein limit of TW gravity is just general relativity. 

Seeing as one of the primary objectives of this paper is to introduce torsion in $\tilde{\Gamma}$, we would like to amend our above definition. The Einstein limit is really the torsionfree Einstein limit, which is a natural weak field limit of torsionfree TW gravity. However, the torsionful Einstein limit will start with the torsionful TW connection and build the identical action (now including torsion), and taking all of $\mathcal{D}, \overline{C}, \mathcal{V}, \mathfrak{u}, \mathfrak{a} \to 0$, as well as $J_0 \to 0$. The motivation for this limit is to recover Einstin-Cartan gravity, wherein $\overline{T}$ is the usual (trace-free part of the) torsion tensor, as in \cite{SpinAndTorsion}. To recover Einstein-Cartan gravity we must take $J_0 \to 0$ to remove the dynamics of $\overline{T}$ endowed through the Gauss-Bonnet term. Further study of the dynamics provided to $\overline{T}$ through the Gauss-Bonnet action may prove insightful and would accurately be viewed as the torsionful Lovelock limit of TW gravity.

\subsection{Electromagnetism on the Thomas Cone}

Given an electromagnetic vector potential $\mathcal{A}_m$ on $M$, we can extend it to live on $VM$ by taking $\mathcal{A}_M = (\mathcal{A}_m + g_m, \frac{1}{\lambda} )$ as in \cite{Brensinger:2020gcv} and described in the introduction. To lift the vector index on $\mathcal{A}_m$ we can actually use any $\kappa_m$ with the transformation law from equation \ref{eq: KappaTransformation}, but we shall continue with the assumption $\kappa_m = g_m$ in this paper. Following the prescription for minimally coupling electrodynamics to gravity, we take 
\begin{align}
    \tilde{\nabla}_M \to \mathscr{D}_M \equiv \tilde{\nabla}_M - i \mathcal{A}_M   
\end{align}
as the gauge covariant derivative with $e$ the coupling constant. Recalling the definition of $\mathcal{K}^A_{\ BCD}$ as 
\begin{align}
    \mathcal{K}^A_{\ BCD} V^B & = [\tilde{\nabla}_C, \tilde{\nabla}_D] V^A + ( \tilde{\Gamma}^B_{\ CD} - \tilde{\Gamma}^B_{\ DC} ) \tilde{\nabla}_B V^A, 
\end{align}
we promote this to 
\begin{align}
    \mathscr{K}^A_{\ \ BCD} V^B & = [ \mathscr{D}_C, \mathscr{D}_D ] V^A + \left( \tilde{\Gamma}^B_{\ CD} - \tilde{\Gamma}^B_{\ DC} \right) \mathscr{D}_B V^A \nonumber \\ 
    & = \mathcal{K}^A_{\ BCD} V^B - i \left( \tilde{\nabla}_C \mathcal{A}_D - \tilde{\nabla}_D \mathcal{A}_C \right) V^A, 
\end{align}
so the full curvature tensor can be succinctly expressed as  
\begin{align}
\label{eq: Definition of Crazy K Maxwell}
    \mathscr{K}^A_{\ \ BCD} = \mathcal{K}^A_{\ BCD} - i \delta^A_{\ B} ( \tilde{\nabla}_C \mathcal{A}_D - \tilde{\nabla}_D \mathcal{A}_C ). 
\end{align}

Before writing the final form of our action, a notational interlude is appropriate. Following \cite{Brensinger:2020gcv}, we introduce the Gauss-Bonnet operator
\begin{align}
\label{eq: GaussBonnetOperator}
    \mathscr{G}_{AE}^{\ \ \ BFCGDH} & = G_{AE} G^{BF} G^{CG} G^{DH} - 4 \delta^A_{\ C} \delta^E_{\ G} G^{BF} G^{DH} + \delta^A_{\ C} \delta^E_{\ G} G^{BD} G^{FH}.
\end{align}
There is no special significance to the operator $\mathscr{G}$ other than aesthetic appeal and computational convenience. Indeed, the above definition allows us to write 
\begin{align}
    \mathscr{G}_{AE}^{\ \ \ BFCGDH} \mathcal{K}^A_{\ BCD} \mathcal{K}^E_{\ FGH} & = \mathcal{K}^{A}_{\ BCD} \mathcal{K}_{A}^{\ BCD} - 4 \mathcal{K}_{AB} \mathcal{K}^{AB} + \mathcal{K}^2, 
\end{align}
a recombination of terms that is exceedingly helpful when deriving field equations. 

Now that we have working definitions of $\mathscr{K}$ and $\mathscr{G}$ in place, we can construct the Maxwell-TW action. It is  
\begin{align} 
    \label{eq: TW Maxwell Action} S_{\text{MTW}} & = \frac{1}{2 \kappa_0} \int \sqrt{|G|} \mathscr{K} d \lambda d^d x + J_0 c \int \sqrt{|G|} \, \left( \mathscr{G}_{AE}^{\ \ BFCGDH} \mathscr{K}^{A}_{\ BCD} \mathscr{K}^{E}_{\ FGH} \right) d \lambda d^d x, 
\end{align}
in which there are again essentially three constituent terms: $\mathscr{K}^A_{\ \ BCD}, \mathscr{K}_{BD}$, and $\mathscr{K}$. We examine their properties in the next several sections.

\subsection{Torsionfree Maxwell-TW Decoupling}

While beginning our Coulombic endeavors let us temporarily revert to assuming $\tilde{\Gamma}$ is torsionfree. From the definition of $\mathscr{K}$ in equation \ref{eq: Definition of Crazy K Maxwell}, we see only the latter portion involves $\mathcal{A}$. Expanding this term, we find 
\begin{align}
    \tilde{\nabla}_C \mathcal{A}_D - \tilde{\nabla}_D \mathcal{A}_C & = \partial_C \mathcal{A}_D - \tilde{\Gamma}^E_{\ DC} \mathcal{A}_E - \partial_D \mathcal{A}_C + \tilde{\Gamma}^E_{\ DC} \mathcal{A}_E = \partial_C \mathcal{A}_D - \partial_D \mathcal{A}_C,  
\end{align}
so for a torsionfree $TW$ connection this resembles the usual electromagnetic tensor. Recognizing this, we write
\begin{align}
    \mathscr{F}_{CD} & = \partial_C \mathcal{A}_D - \partial_D \mathcal{A}_C,   
\end{align}
which in block form is 
\begin{align}
\mathscr{F}_{GH} & =  
    \left( \begin{array}{@{}c|c@{}}
    \partial_g \mathcal{A}_h - \partial_h \mathcal{A}_g & \, 0 \, \\ \hline
    0 & \, 0 \,
  \end{array} \right)_{GH} \, .
\end{align}
Clearly $\mathscr{F}$ is the just usual electromagnetic tensor when restricted to $M$. 

We now wish to examine the relevant terms in the Maxwell-TW action to understand possible coupling between the potential $\mathcal{A}$ and $\tilde{\Gamma}$. First, the projective Riemann contribution is 
\begin{align}
\label{eq: Maxwell Crazy K torsionfree Riemann tensor}
    \mathscr{K}^A_{\ \ BCD} \mathscr{K}_A^{\ BCD} & = \mathcal{K}^A_{\ BCD} \mathcal{K}_A^{\ BCD} - 2 i \mathcal{K}^A_{\ BCD} \delta_A^{\ B} \mathscr{F}^{CD} - (d+1) \mathscr{F}_{CD} \mathscr{F}^{CD} \nonumber \\ 
    & = \mathcal{K}^A_{\ BCD} \mathcal{K}_A^{\ BCD} - (d+1) \mathscr{F}_{CD} \mathscr{F}^{CD},  
\end{align}
where the middle term $\mathcal{K}^A_{\ ACD} \mathscr{F}^{CD}$ vanishes. In the projective Ricci term we obtain 
\begin{align}
    \mathscr{K}_{BD} \mathscr{K}^{BD} & = \mathcal{K}_{BD} \mathcal{K}^{BD} - \mathscr{F}_{BD} \mathscr{F}^{BD}, 
\end{align} 
and the projective Einstein-Hilbert term is 
\begin{align}
    \mathscr{K} & = \mathcal{K}, \qquad \mathscr{K}^2 = \mathcal{K}^2, 
\end{align}
where the scalar curvature $\mathcal{K}$ is unchanged under minimal coupling. Therefore, putting our three components together tells us that $\mathcal{A}$ completely decouples from $\mathcal{D}$ and $\overline{C}$ in the Maxwell-TW Lagrangian.

\subsection{Generalized Maxwell-TW Decoupling}

It is well known that including torsion into a gravitational theory can induce gravitational coupling to spinors \cite{SpinAndTorsion, DiracEquationTorsionZecca} and gauge fields \cite{CartanYMMatter}. Now that we have seen how a $U(1)$ gauge field acts on $VM$, we wish to understand ramifications of including torsion. Again for minimal coupling we have   
\begin{align}
\label{eq: Crazy K Maxwell torsion}
    \mathscr{K}^A_{\ BCD} = \mathcal{K}^A_{\ BCD} - i \delta^A_{\ B} ( \partial_C \mathcal{A}_D - \partial_D \mathcal{A}_C ), 
\end{align}
and recognizing the latter term as the electromagnetic field strength we again define 
\begin{align}
    \mathscr{F}_{CD}& = \partial_C \mathcal{A}_D - \partial_D \mathcal{A}_C.  
\end{align}
In block form this is 
\begin{align}
\mathscr{F}_{GH} & = 
    \left( \begin{array}{@{}c|c@{}}
   \partial_g \mathcal{A}_h - \partial_h \mathcal{A}_g & 0 \\ \hline
    0 & 0 
    \end{array} \right)_{GH} \, .
\end{align}
Equation \ref{eq: Crazy K Maxwell torsion} tells us the projective Riemann contribution is 
\begin{align}
\label{eq: Maxwell Crazy K torsion Riemann tensor}
    \mathscr{K}^A_{\ BCD} \mathscr{K}_A^{\ BCD} & = \mathcal{K}^A_{\ BCD} \mathcal{K}_A^{\ BCD} - 2 i \mathcal{K}_A^{\ ACD} \mathscr{F}_{CD} - (d+1) \mathscr{F}_{CD} \mathscr{F}^{CD}, 
\end{align}
where the middle term will no longer vanishes due to the torsion fields. The projective Ricci tensor term becomes
\begin{align}
\label{eq: Maxwell Crazy K torsion Ricci tensor}
    \mathscr{K}_{BD} \mathscr{K}^{BD} & = \mathcal{K}_{BD} \mathcal{K}^{BD} - \mathscr{F}_{BD} \mathscr{F}^{BD}, 
\end{align}
and the projective Ricci scalar is 
\begin{align}
\label{eq: Maxwell Crazy K torsion Ricci Scalar}
    \mathscr{K} = \mathcal{K}, \qquad \mathscr{K}^2 = \mathcal{K}^2. 
\end{align}

Examining equations \ref{eq: Maxwell Crazy K torsion Ricci Scalar} and \ref{eq: Maxwell Crazy K torsion Ricci tensor}, we see $\mathscr{F}$ and $\mathcal{K}$ decouple, so the only possible interactions between $\mathcal{A}$ and $\tilde{\Gamma}$ come from the factor of $\mathcal{K}_A^{\ ACD} \mathscr{F}_{CD}$ in equation \ref{eq: Maxwell Crazy K torsion Riemann tensor}. This term can be expanded to  
\begin{align}
\label{eq: Maxwell torsion Riemann cross term}
    \mathcal{K}_A^{\ ACD} \mathscr{F}_{CD} & = \big( \partial_d \mathfrak{a}_c - \partial_c \mathfrak{a}_d - 2 g_d \partial_c \mathfrak{u}\big) \mathscr{F}^{cd},  
\end{align}
where $\mathfrak{u} = \mathfrak{u}^a_{\ a}$. The exact form of equation \ref{eq: Maxwell torsion Riemann cross term} will be quite relevant in some future projects, such as determining field equations or studying quantum interactions. However, for this paper we are not interested in the exact expression, but just the role each field plays. Indeed, two properties are apparent: first, as required, letting the torsion fields vanish means the entire expression vanishes, recovering equation \ref{eq: Maxwell Crazy K torsionfree Riemann tensor} in the weak field limit. Second, $\mathcal{A}$ couples to $\mathfrak{a}$ and $\mathfrak{u}$, which provides interaction terms in the Lagrangian. 

Finally, we include one last observation that is both the most subtle and the most important: $\mathcal{D}$ and $\overline{C}$ are not present in this expression. Results in \cite{Brensinger:2017gtb, Brensinger:2019mnx, Brensinger:2020gcv} (summarized in the introduction) show that $\mathcal{D}$ and $\overline{C}$ act as geometric sources through the energy momentum tensor. From equation \ref{eq: Maxwell torsion Riemann cross term} we see that, even in the presence of torsion fields, $\mathcal{D}$ and $\overline{C}$ $\textit{will}$ contribute to the energy momentum tensor, but will $\textit{not}$ couple to $\mathcal{A}$ in the Maxwell-TW Action! This decoupling is sufficiently remarkable that we would like to characterize it as the Maxwell Miracle. 

The main result from this section is that in the Maxwell-TW action the fields $\mathcal{D}$ and $\overline{C}$, introduced through the TW connection, naturally appear as contributions to an electromagnetically dark sector. One may speculate if this property is unique to the Maxwell-TW action, or if more general Lagrangians share this `miracle property'. Indeed, it may be worth further study to examine the relevant dynamics in the context of, say, including a Born-Infeld action \cite{BornInfeldGravity}, adding contributions from higher order Lovelock terms \cite{Lovelock:1971yv, PalatiniProjective:2017}, or by including a Palatini expansion of the curvature tensors such as in \cite{Borunda:2008}.

\section{Yang-Mills Interactions}

\label{sec: YM Stuff}

\subsection{Nonabelian Gauge Fields on the Thomas Cone}

We would now like to extend our above results to the context of a non-abelian gauge theory and describe the mechanism by which $\mathcal{D}$ and $\overline{C}$ decouple from general Yang-Mills fields. After the previous section there will be few surprises, but it is nevertheless insightful to detail the precise construction of nonabelian gauge fields on the Thomas cone, especially regarding extensions of Lie algebra representations from $M$ to $VM$. It is also likely that, despite the classical similarities, moving to a nonabelian gauge group will have significant quantum repercussions and thus warrant a separate treatment. 

Our immediate priority is establishing conventions. Let $\mathfrak{g}$ be any compact semisimple Lie algebra and $G$ a Lie group with Lie algebra $\mathfrak{g}$. We really have $G = SU(N)$ in mind as in Yang-Mills, but all results in this section are valid for any compact semisimple $\mathfrak{g}$. In particular, taking $\mathfrak{g} = \mathfrak{su} (3) \times \mathfrak{su} (2) \times \mathfrak{u} (1)$, we may consider the Standard Model. 

Let $T^{\alpha}$ be the generators of $\mathfrak{g}$ in the adjoint representation, where $\alpha$ indexes the basis elements. When working with Yang-Mills connections we generally display the Greek basis indices of $\mathfrak{g}$ but suppress the internal representation indices of $T^{\alpha}$. It is a minor travesty that our convention inexorably results in Latin spacetime indices and Greek group representation indices, but the benefits of the Mera convention ultimately outweigh the initial notational translation. Finally, the structure constants are defined by the usual commutation relation $[ T^{\alpha}, T^{\beta} ] = i f^{\alpha \beta}_{\ \ \ \gamma} T^{\gamma}$. By assumption that $\mathfrak{g}$ is compact semisimple we can take the structure constants to be totally antisymmetric; this is not always true for arbitrary Lie algebras. Of course, $\lambda$ will not be used for a Lie algebra index and is reserved for the coordinate on $VM$. 

The Yang-Mills potential on $M$ is the usual sum $\mathcal{A}_{m \alpha} T^{\alpha}$, which is often described as a $\mathfrak{g}$-valued $1$-form. By no means do we wish to contest this perspective, but here it will prove more convenient to view $\mathcal{A}_{m \alpha}$ as a collection of vectors indexed by $\alpha$. For each fixed $\alpha$ we can extend $\mathcal{A}_{m \alpha}$ to $VM$ by the usual lifting, i.e., 
\begin{align}
     \mathcal{A}_{M \alpha} = \left( \mathcal{A}_{m \alpha} + g_m, \tfrac{1}{\lambda} \right), 
\end{align}
which is a well-formed object for any particular value of $\alpha$. As by now expected, the appearance of $g_m$ is the result of the Roberts gauge, and a more general form would include an arbitrary choice of $\kappa_m$. This is planned for future study. 

To have a genuine Yang-Mills potential we need a consistent way to sum our collection of vectors over the generators $T^{\alpha}$. To this end, note that both $g_m$ and $\tfrac{1}{\lambda}$ commute with each of the $A_{m \alpha}$, so these terms should only appear in the center of $\mathfrak{g}$. This requires the potential $\mathscr{A}$ to take the form \cite{Brensinger:2020gcv}
\begin{align}
\label{eq: YM Potential on VM}
     \mathscr{A}_M \equiv \left( \mathcal{A}_{m \alpha} T^{\alpha} + g_m I, \tfrac{1}{\lambda} I \right), 
\end{align}
where $I$ is the identity of the same dimension as the $T^{\alpha}$. To compute the curvature we decompose $\mathscr{A}$ as 
\begin{align}
     \mathscr{A}_M = \left( \mathcal{A}_{m \alpha} T^{\alpha}, 0 \right) + \left( g_m, \tfrac{1}{\lambda} \right) I = \left( \mathcal{A}_{m \alpha} T^{\alpha}, 0 \right) + g_M I, 
\end{align}
which also demonstrates that $\mathscr{A}$ is tensorial on $VM$. With this in place, we can define the gauge covariant derivative 
\begin{align}
    \mathscr{D}_M = \tilde{\nabla}_M - i \mathscr{A}_M. 
\end{align}
The curvature is again (see appendix) 
\begin{align}
    \mathscr{K}^A_{\ \ BCD} V^B & = [ \mathscr{D}_C, \mathscr{D}_D ] V^A + ( \tilde{\Gamma}^B_{\ CD} - \tilde{\Gamma}^B_{\ DC} ) \mathscr{D}_B V^A, 
\end{align}
which boils down to 
\begin{align}
\label{eq: Definition of Crazy K YM no F}
    \mathscr{K}^A_{\ \ BCD} = \mathcal{K}^A_{\ BCD} - i e \delta^A_{\ B} \left( \tilde{\nabla}_{[C} \mathscr{A}_{D]} - i \mathscr{A}_{[C} \mathscr{A}_{D]} \right). 
\end{align} 
As in the case of electromagnetism, the latter term is the field strength of $\mathscr{A}$, so we can succinctly write  
\begin{align}
\label{eq: Definition of Crazy K YM with F}
    \mathscr{K}^A_{\ \ BCD} = \mathcal{K}^A_{\ BCD} - i \delta^A_{\ B} \mathscr{F}_{CD}, 
\end{align} 
where we have introduced 
\begin{align}
\label{eq: Yang-Mills Crazy F Definition}
    \mathscr{F}_{CD} = \tilde{\nabla}_C \mathscr{A}_D - \tilde{\nabla}_D \mathscr{A}_C - i \mathscr{A}_{[C} \mathscr{A}_{D]}. 
\end{align} 
The Yang-Mills-TW action is then 
\begin{align} 
    \label{eq: YM-TW Action} S_{\text{YMTW}} & = \tfrac{1}{2 \kappa_0} \int \sqrt{|G|} \mathscr{K} d \lambda d^d x + J_0 c \int \sqrt{|G|} \mathscr{G}_{AE}^{\ \ BFCGDH} \mathscr{K}^{A}_{\ \ BCD} \mathscr{K}^{E}_{\ \ FGH} d \lambda d^d x, 
\end{align}
which is of the same form as in equation \ref{eq: TW Maxwell Action}.

\subsection{Torsionfree Yang-Mills-TW Decoupling}

To classify coupling between the Yang-Mills potential $\mathscr{A}$ and $\tilde{\Gamma}$ we again consider the torsionfree case prior to full generality. Assuming $\tilde{\Gamma}$ is torsionfree, we can use equation \ref{eq: Yang-Mills Crazy F Definition} to expand $\mathscr{F}$ in block form as 
\begin{align}
\mathscr{F}_{GH} & = 
    \left( \begin{array}{@{}c|c@{}}
   \partial_g \mathcal{A}_h - \partial_h \mathcal{A}_g + f^{\alpha \beta}_{\ \ \gamma} T^{\gamma} \mathcal{A}_{g \alpha} \mathcal{A}_{h \beta} & 0 \\ \hline
    0 & 0 
    \end{array} \right)_{GH} \, ,
\end{align}
which is the usual Yang-Mills field strength in the first block $M$. 

Again we wish to examine the coupling between $\mathscr{F}$ and the TW sector. To this end, we first find the Riemann tensor, 
\begin{align}
\label{eq: YM Crazy K torsionfree Riemann tensor}
    \mathscr{K}^A_{\ \ BCD} \mathscr{K}_A^{\ BCD} & = \mathcal{K}^A_{\ BCD} \mathcal{K}_A^{\ BCD} - 2 i \mathcal{K}^A_{\ BCD} \delta_A^{\ B} \mathscr{F}^{CD} - (d+1) \mathscr{F}_{CD} \mathscr{F}^{CD} \nonumber \\ 
    & = \mathcal{K}^A_{\ BCD} \mathcal{K}_A^{\ BCD} - (d+1) \mathscr{F}_{CD} \mathscr{F}^{CD}.  
\end{align}
As by now expected, the term $\mathcal{K}^A_{\ ACD} \mathscr{F}^{CD}$ vanishes identically in the above expression. Taking the trace gives us the Ricci term, 
\begin{align}
    \mathscr{K}_{BD} \mathscr{K}^{BD} & = \mathcal{K}_{BD} \mathcal{K}^{BD} - \mathscr{F}_{BD} \mathscr{F}^{BD}, 
\end{align}
and another trace gives the Ricci scalar, 
\begin{align}
    \mathscr{K} & = \mathcal{K}, \qquad \mathscr{K}^2 = \mathcal{K}^2. 
\end{align}
As before, no new dynamics are introduced through the Ricci scalar. Putting all of this together implies a total decoupling between Yang-Mills fields $\mathcal{A}$ and TW fields $\mathcal{D}$ and $\overline{C}$ in the context of the torsionfree Yang-Mills-TW Lagrangian, in exact analogue with the electromagnetic case.

\subsection{Generalized Yang-Mills-TW Decoupling}

The final case to explore in our study of decoupling is the Yang-Mills action with torsion, which is the natural subsummation of the last two sections. Including torsion in the TW connection still leaves the block form of $\mathscr{F}$ as
\begin{align}
\mathscr{F}_{GH} & = 
    \left( \begin{array}{@{}c|c@{}}
   \partial_g \mathcal{A}_h - \partial_h \mathcal{A}_g + f^{\alpha \beta}_{\ \ \gamma} T^{\gamma} \mathcal{A}_{g \alpha} \mathcal{A}_{h \beta} & 0 \\ \hline
    0 & 0 
    \end{array} \right)_{GH} \, .
\end{align}

We find that the projective Riemann squared contribution is 
\begin{align}
\label{eq: YM Crazy K torsion Riemann tensor}
    \mathscr{K}^A_{\ BCD} \mathscr{K}_A^{\ BCD} & = \mathcal{K}^A_{\ BCD} \mathcal{K}_A^{\ BCD} - 2 i \mathcal{K}_A^{\ ACD} \mathscr{F}_{CD} - (d+1) \mathscr{F}_{CD} \mathscr{F}^{CD}, 
\end{align}
the projective Ricci tensor squared term becomes
\begin{align}
\label{eq: YM Crazy K torsion Ricci tensor}
    \mathscr{K}_{BD} \mathscr{K}^{BD} & = \mathcal{K}_{BD} \mathcal{K}^{BD} - \mathscr{F}_{BD} \mathscr{F}^{BD}, 
\end{align}
and the projective Ricci scalar is 
\begin{align}
\label{eq: YM Crazy K torsion Ricci Scalar}
    \mathscr{K} = \mathcal{K}, \qquad \mathscr{K}^2 = \mathcal{K}^2. 
\end{align}
As is the case in the Maxwell coupling, the mechanism by which the Yang-Mills fields could potentially couple to $\mathcal{D}$ and $\overline{C}$ in the TW Lagrangian is from the presence of the term 
\begin{align}
\mathcal{K}_A^{\ ACD} \mathscr{F}_{CD}.  
\end{align}
Expanding this out, we find 
\begin{align}
\label{eq: YM torsion Riemann cross term}
    \mathcal{K}_A^{\ ACD} \mathscr{F}_{CD} & = \left( \partial_d \mathfrak{a}_c - \partial_c \mathfrak{a}_d - 2 g_d \partial_c \mathfrak{u} \right) \mathscr{F}^{cd}. 
\end{align}
On paper this term is identical to the $U(1)$ case, the only difference being in the components of $\mathscr{F}$. We see that $\mathcal{D}$ and $\overline{C}$ decouple from any Yang-Mills field, even in the presence of torsion, which we would like to analogously dub the Yang-Mills Miracle. 

Before concluding this section we offer several remarks. First, in some sense the Maxwell and Yang-Mills miracles are poorly named as the decoupling is not due to any property of $SU(N)$ gauge theories! Instead they are really special cases of a `Projective Miracle,' as the decoupling comes from the form of $\mathcal{K}^A_{\ \, BCD}$. Indeed, the form of the term in equation \ref{eq: YM torsion Riemann cross term} is identical for any two-form in the place of $\mathscr{F}$, underpining the assumption that $\mathfrak{g}$ is compact and semisimple: any compact semisimple $\mathfrak{g}$ has totally antisymmetric structure constants $f^{\alpha \beta}_{\ \ \gamma}$ in the adjoint representation, implying that $\mathscr{F}$ is antisymmetric, and therefore does not interact with $\mathcal{D}$ or $\overline{C}$. It may be interesting to investigate the possibility that more general Lie algebras induce such coupling. 

Presently much remains before detailed study of the quantum theory. Although we cannot hope to establish the quantum theory here, several prior formulae provide insights into conceivable results. One we wish to highlight is a manifestation of tree level diagrams. In equation \ref{eq: YM torsion Riemann cross term} there is a decoupling of $\mathcal{A}$ from the TW fields, contrasting with the fact that $\mathfrak{a}$ and $\mathfrak{u}$ couple to $\mathscr{A}$ as well as $\mathcal{D}$ and $\overline{C}$. It is therefore possible that in multiple vertex diagrams the torsion fields allow interactions between standard model fields and fields in the dark sector. This may conceivably be observed as an interaction beginning with baryonic matter that produces particles in the dark sector, especially if the torsion fields decay rapidly. Despite a likely first impression, this possibility does not rule out phenomological viability; each vertex of this interaction has a factor of $J_0 c$. Recent work \cite{InflationNewPaper:2022} gives an order-of-magnitude estimate for $J_0$ in a dark energy dominated universe at $10^{10} \alpha \hbar$ ($\alpha$ is the fine structure constant), so the interaction term is extremely small. The authors of \cite{InflationNewPaper:2022} also found other situations for which $J_0$ can be much larger, but the dark energy solution is the main interest to us. With $J_0$ taking a small value, it may then be reasonable to speculate that at sufficiently high energies there is a thermal exchange between TW and GUT fields. Given the energy scales, this suggests the possibility that $\mathcal{D}$ and $\overline{C}$ were prominent in the very early universe but have since decoupled from baryonic matter.

\section{Thomas-Whitehead-Yang-Mills-Dirac Action}

\subsection{Lifting Geometric Actions}

\subsubsection{Lifting  Fundamental Representations}

This section is the capstone to this paper, synthesizing all previous constructions with spinors, projective geometry, torsion, and Yang-Mills fields. All of the actions hitherto discussed will be combined to form the overall action, which is the sum of the Thomas-Whitehead, Yang-Mills, and Dirac actions. By now we have all of the necessary background in place, with the minor exception that we will also need the lift of the fundamental representation of gauge groups to $VM$ when coupling $\mathcal{A}$ to fermions. 

Again let $\mathfrak{g}$ be a compact semisimple Lie algebra, $G$ a Lie group with Lie algebra $\mathfrak{g}$, and $T^{\alpha}$ be the generators of $\mathfrak{g}$ in the fundamental representation, where $\alpha$ indexes the basis elements. The structure constants satisfy $[ T^{\alpha}, T^{\beta} ] = i f^{\alpha \beta}_{\ \ \ \gamma} T^{\gamma}$ and are totally antisymmetric. Again following \cite{Brensinger:2020gcv}, the lift of $\mathscr{A}$ to $VM$ is formally identical to that in equation \ref{eq: YM Potential on VM}, i.e. 
\begin{align}
     \mathscr{A}_M \equiv \left( \mathcal{A}_{m \alpha} T^{\alpha} + g_m I, \tfrac{1}{\lambda} I \right), 
\end{align}
where $I$ has the same dimension as the $T^{\alpha}$. The only change in this construction from the lift of $\mathscr{A}$ in the adjoint representation is the internal indices, which have no interaction with the actual lifting mechanism.

\subsubsection{Lifting Yang-Mills Actions}

Now that we have established the form of the Yang-Mills field strength on $VM$, we can form the Yang-Mills action by the usual quadratic term. Accounting for the $\lambda$ dependence, we have 
\begin{align}
    S_{YM} & = \frac{1}{e^2} \int \sqrt{|G|} \mathscr{F}_{AB} \mathscr{F}^{AB} d \lambda d^d x = \frac{1}{e^2} \left( \int \frac{\lambda_0}{\lambda} d \lambda \right) \int \sqrt{|g|} F_{ab} F^{ab} d^d x 
\end{align}
where $F_{ab}$ is the usual Yang-Mills tensor on $M$. The factor of $\lambda_0$ again arises from equation \ref{eq:Relation Between Metric Determinants} and ensures the overall action is dimensionless. It can readily be observed that the equations of motion from the Yang-Mills action are identical to the standard equations in a classical gauge theory, up to the integral of $\lambda$ in front, which, as suggested in \cite{Brensinger:2020gcv}, seems to be acting as a renormalization factor. Further work on this is under way.

\subsubsection{Lifting Dirac Actions}

The Dirac action on $VM$ is
\begin{align}
        S_{\text{Dirac}} & = \int \sqrt{|G|} \, \overline{\Psi} \left( i \tilde{\slashed{\nabla}} - m \right) \Psi d \lambda d^d x. 
\end{align}
Unraveling the definition of $\tilde{\slashed{\nabla}}$ from equation \ref{eq: unexpanded baby Dirac operator}, we minimally couple the Dirac operator to our Yang-Mills field by the usual concoction 
\begin{align}
    \tilde{\slashed{\nabla}} \to \slashed{\mathscr{D}} = \tilde{\slashed{\nabla}} - i \slashed{\mathscr{A}}.
\end{align}
Then expanding the minimally coupled Dirac action, we obtain 
\begin{align}
    S_{\text{Dirac}} & = \int \sqrt{|G|} \, \overline{\Psi} \bigg( i \slashed{\partial} + \slashed{\mathscr{A}} + i \tilde{\gamma}^{\lambda} \left( \Xi_{\lambda} + \tfrac{i}{\lambda} \right) \nonumber \\ 
        & \qquad + \tfrac{i}{2(d+1)} \left( \left( \slashed{\partial} e^n_{\ \underline{b}} \right) \gamma_n^{\ \underline{b}} + \mathcal{O} - 2 i \left( \chi + \tfrac{d - \mathfrak{u}}{\lambda_0} \right) \gamma^{\underline{d+1}} \right) - m \bigg) \Psi d \lambda d^d x, 
\end{align}
where each of the fields are now explicit. As when originally constructing $\tilde{\slashed{\nabla}}$, we let $\slashed{\partial} = \gamma^m \partial_m$ and introduced the fields 
\begin{align}
    \mathcal{O} & = \gamma^m \left( \Pi_{nmp} - g_p \left( g_{nm} - \mathfrak{u}_{nm} \right) \right) \left( \gamma^n \gamma^p - \gamma^p \gamma^n \right) - \weight \gamma^m g_m,  \nonumber \\ 
    \chi & = \lambda_0 \left( g_n \Pi^n_{\ mp} - g_p \left( g_m - g_n \mathfrak{u}^n_{\ m} - \mathfrak{a}_m \right) +  \partial_m g_p - \mathcal{D}_{mp} + \mathcal{V}_{mp} \right) \gamma^m \gamma^p, \quad \text{and} \nonumber \\ 
    \Xi_{\lambda} & = \partial_{\lambda} - \tfrac{\lambda_0}{\lambda (d+1)} \left( g_n \mathfrak{u}^n_{\ p} + \mathfrak{a}_p \right) \gamma^p \gamma^{\underline{d+1}}.
\end{align}
Noticing the Yang-Mills coupling constant appears with a factor of $\tfrac{1}{\lambda}$, it is reasonable to speculate that the $\lambda$ component may play some role in renormalization. This is another avenue of further research that we hope to explore in the future.

\subsection{The Holy Trinity}

Now that we have an understanding of all of the relevant geometries on the Thomas cone, we can construct an action simultaneously encompassing the TW, Yang-Mills, and Dirac sectors that is both projectively invariant and covariant. The overall action is 
\begin{align}
    S = S_{\text{TW}} + S_{\text{YM}} + S_{\text{Dirac}},
\end{align}
where the respective components are 
\begin{align}
    S_{\text{TW}} & = \frac{1}{2 \kappa_0} \int \sqrt{|G|} \, \mathscr{K} d \lambda d^d x \scriptscriptstyle{+} J_0 c \int \sqrt{|G|} \mathscr{G}_{AE}^{\ \ \ BFCGDH} \mathscr{K}^{A}_{\ BCD} \mathscr{K}_{A}^{\ BCD} d \lambda d^d x, \nonumber \\ 
    S_{\text{YM}} & = \frac{1}{e^2} \int \sqrt{|G|} \, \mathscr{F}_{AB} \mathscr{F}^{AB} d \lambda d^d x, \quad \text{and} \nonumber \\ 
    S_{\text{Dirac}} & = \int \sqrt{|G|} \, \overline{\Psi} \left( i \slashed{\mathscr{D}} - m \right) \Psi d \lambda d^d x. 
\end{align}
Written under one integral sign, the total action is 
\begin{align}
\label{eq: Total Action Compressed}
    S & = \int \sqrt{|G|} \Big( \overline{\Psi} \left( i \slashed{\mathscr{D}} - m \right) \Psi + \tfrac{1}{e^2} \mathscr{F}_{AB} \mathscr{F}^{AB} \nonumber \\ 
        & \qquad + \tfrac{1}{2 \kappa_0} \mathscr{K} + J_0 c \, \mathscr{G}_{AE}^{\ \, BFCGDH} \mathscr{K}^{A}_{\ BCD} \mathscr{K}^{E}_{\ FGH} \Big) d \lambda d^d x. 
\end{align}
Although equation \ref{eq: Total Action Compressed} offers a clean expression of all of the material covered in this paper, it leaves much of the underlying machinery regarding field interactions hidden. For the purpose of completeness we have included the expansion in section \ref{sec: The Complete Action} in the appendix.

\section{Conclusion}

The geometric structure associated with the volume bundle naturally includes a gauge symmetry over spacetime. We have seen how this symmetry arises and the impact it has on lifting tensor and spinor fields from $M$ to $VM$. It is now clear that all work thus far on TW gravity has been constrained to the Roberts gauge, a particular gauge fixing condition. This geometric lifting also induces a weight on spinors which ultimately contributes to a mass term. 

Although TW gravity is naturally intertwined with the Palatini formalism, the effect of torsion on gravitational interactions has not previously been studied in this context. By including torsion in the TW connection we have seen that the torsion fields add contributions to the Dirac operator through both ordinary and chiral mass terms. Further, we also saw that the torsion fields appear in the Yang-Mills sector, interacting with both Yang-Mills and TW fields in $n$-point correlation functions suppressed by factors of $J_0 c$. Given the exceedingly small scale of $J_0 c$, this raises the possibility that at extreme energies the Yang-Mills and TW fields may have a thermal exchange mediated by torsion. In particular, this may have played a role in the very early universe. 

In addition to understanding these interaction terms, we have constructed the lifting of Yang-Mills and Dirac actions to $VM$, allowing the construction of the total action $S = S_{\text{TW}} + S_{\text{YM}} + S_{\text{D}}$. With this action in place there is now a natural route forward to computing field equations and to begin quantizing TW gravity. 

A natural question is how moving away from the Roberts gauge may affect the dynamics of TW gravity. At the fundamental level it is clear this will impact the process of lifting tensor and spinor fields from $M$ to $VM$, but how this will manifest in relevant actions and field equations is currently nebulous. This is presently under investigation.

\section{Acknowledgments}

The authors would like to thank the Diffeomorphisms and Geometry Research Group at the University of Iowa for constructive conversations throughout this work. In particular, the authors would like to thank Vincent Rodgers for his insight and wisdom. We would also like to thank Michael Connolly, Tyler Grover, Connor Lindsay, Salvatore Quaid, and Kory Stiffler for stimulating discussions that provided valuable contributions.

\appendix

\section{Assumptions, Conventions, and Useful Identities}

\subsection{Assumptions and Conventions}

Everywhere in this paper we assume all manifolds, functions, tensors, spinors, etc. are smooth (infinitely differentiable), and that our spacetime manifold $M$ is Lorentizian and orientable. Unless otherwise stated we make no further assumptions. Our work employs the language of abstract index notation as there exist no circumstances in which coordinate free notation provides any advantage over the use of indices. As usual repeated indices indicate summation, and parentheses and brackets denote symmetrization and antisymmetrization, respectively. For example, if $T_{ab}$ is a rank 2 tensor we have 
\begin{align}
    T_{(ab)} = T_{ab} + T_{ba}, \qquad T_{[ab]} = T_{ab} - T_{ba}, 
\end{align}
where we do \emph{not} include a factor of $\frac{1}{2}$. On both the manifold $M$ and the bundle $VM$ the usual index juggling patterns apply. It is important to distinguish indices between $M$ and $VM$, which is elegantly accomplished with the Mera convention. 

The chosen signature is mostly plus, so $\eta_{ab} = \text{diag} (-1,1,1,1)$. Furthermore, the $\lambda$ component has a spacelike signature, so $\eta_{\underline{A} \underline{B}} = \text{diag} (-1,1,1,1,1)$. An immediate corollary of this is the requirement for a minus sign in the definition of the gamma matrices; indeed, the defining relations are 
\begin{align}
    \{ \gamma^a, \gamma^b \} = - 2 g^{ab}, \qquad \{ \tilde{\gamma}^A, \tilde{\gamma}^B \} = - 2 G^{AB}, 
\end{align}
for the algebras on $M$ and $VM$, respectively.

\subsection{Definitions, Summarized}

The piecemeal parts used to construct fields are 
\begin{align}
    g_a = - \tfrac{1}{d+1} \partial_a \log \sqrt{|g|}, \qquad j_a = \partial_a \log{J^{-\frac{1}{d+1}}},  
\end{align}
and the metric and its inverse on $VM$ are 
\begin{align} 
G_{AB} = 
    \left(\begin{array}{@{}c|c@{}}
    g_{ab} + \lambda_0^2 g_a g_b & \frac{\lambda_0^2}{\lambda} g_a \\ \hline
    \frac{\lambda_0^2}{\lambda} g_b & \frac{\lambda_0^2}{\lambda^2}
  \end{array} \right)_{AB} \, , \qquad G^{AB} = 
  \left(\begin{array}{@{}c|c@{}}
    g^{ab} & - \lambda g^{am} g_m \\ \hline
    - \lambda g^{bm}g_m & \frac{\lambda^2}{\lambda_0^2} + \lambda^2 g^{mn} g_m g_n 
  \end{array} \right)^{AB} \, .  
\end{align}
Their determinants satisfy the relations  
\begin{equation} 
    |G| = \frac{\lambda^2_0}{\lambda^2} \; |g|, \qquad  \sqrt{|G|} = \frac{\lambda_0}{\lambda} \sqrt{|g|}.  
\end{equation}
The fundamental projective invariant with torsion is 
\begin{align}
    \Pi^a_{\ bc} & = \Gamma^a_{\ bc} - \tfrac{1}{d+1} \left( \tfrac{d}{d-1} \Gamma^e_{\ ec} - \tfrac{1}{d-1} \Gamma^e_{\ ce} \right) \delta^a_{\ b} - \tfrac{1}{d+1} \left( \tfrac{d}{d-1} \Gamma^e_{\ be} - \tfrac{1}{d-1} \Gamma^e_{\ eb} \right) \delta^a_{\ c},   
\end{align}
which decomposes as 
\begin{align}
    \Pi^a_{\ bc} & = \hat{\Pi}^a_{\ ab} + \overline{C}^a_{\ bc} + \overline{T}^a_{\ bc}, 
\end{align}
where 
\begin{align}
    \hat{\Pi}^a_{\ bc} & = \hat{\Gamma}^a_{\ bc} - \tfrac{1}{d+1} \delta^a_{\ b} \hat{\Gamma}^e_{\ ec} - \tfrac{1}{d+1} \delta^a_{\ c} \hat{\Gamma}^e_{\ eb}, \nonumber \\ 
    \overline{C}^a_{\ bc} & = C^a_{\ bc} - \tfrac{1}{d+1} \delta^a_{\ b} C^e_{\ ec} - \tfrac{1}{d+1} \delta^a_{\ c} C^e_{\ eb}, \nonumber \\ 
    \overline{T}^a_{\ bc} & = T^a_{\ bc} - \tfrac{1}{d+1} \delta^a_{\ b} T^e_{\ ec} - \tfrac{1}{d+1} \delta^a_{\ c} T^e_{\ eb}.  
\end{align}

The projective Riemann curvature tensor is defined through a commutator of covariant derivatives,
\begin{align}
    \mathcal{K}^A_{\ BCD} V^B & = [\tilde{\nabla}_C, \tilde{\nabla}_D] V^A + ( \tilde{\Gamma}^B_{\ CD} - \tilde{\Gamma}^B_{\ DC} ) \tilde{\nabla}_B V^A, 
\end{align}
and the projective Ricci tensor is the contraction over the first and third indices. Ricci scalars are the trace of their associated Ricci tensors: 
\begin{align}
    \mathcal{K}_{BD} & = \delta^C_{\ A} \mathcal{K}^A_{\ BCD}, \qquad 
    \mathcal{K} = G^{BD} \mathcal{K}_{BD}. 
\end{align}
The tensor $\mathcal{K}^A_{\ BCD}$ houses the equi-projective Riemann tensor as a component, which is defined as 
\begin{align}
    \mathcal{R}^a_{\ bcd} & = \partial_c \Pi^a_{\ db} - \partial_d \Pi^a_{\ cb} + \Pi^a_{\ ce} \Pi^e_{\ db} -  \Pi^a_{\ de} \Pi^e_{\ cb}. 
\end{align}
As before, the equi-projective Ricci tensor is the trace over the first and third index, and the equi-projective Ricci scalar is the trace of the equi-projective Ricci tensor:
\begin{align}
    \mathcal{R}_{bd} & = \delta^c_{\ a} \mathcal{R}^a_{\ bcd}, \qquad 
    \mathcal{R} = g^{bd} \mathcal{R}_{bd}. 
\end{align}
All of the aforementioned  equi-projective objects are invariant under projective transformations, but are not tensorial under ordinary coordinate transformations. 

Gamma matrices on $M$ take the usual defining relation 
\begin{align}
    \{ \gamma^a, \gamma^b \} = - 2 g^{ab}.
\end{align}
When using gamma matrices we assume $M$ has even dimension. In this case the product 
\begin{align}
    \gamma^{d+1} = i^{\lfloor \frac{d}{2} \rfloor + 1} \gamma^0 \gamma^1 \dots \gamma^{d-1} 
\end{align}
acts as a chirality matrix while simultaneously extending the representation by one dimension. It satisfies 
\begin{align}
    \{ \gamma^{d+1} , \gamma^m \} & = 0, \qquad 
    (\gamma^{d+1})^2 = I_{\mathfrak{D}}.
\end{align}
This is not to be confused with the algebra of gamma matrices on $VM$, which satisfy 
\begin{align}
    \{ \tilde{\gamma}^M, \tilde{\gamma}^N \} = - 2 G^{MN} I_{\mathfrak{D}}, 
\end{align}
a condition that the chirally extended $\gamma$ do not. The gamms matrices on $VM$ take the explicit form 
\begin{align}
    \tilde{\gamma}^n & = \gamma^n, \qquad 
    \tilde{\gamma}^{\lambda} = \tfrac{\lambda}{\lambda_0} ( i \gamma^{d+1} - \lambda_0 g_m \gamma^m ) = \tfrac{\lambda}{\lambda_0} ( \gamma^{\underline{d+1}} - \lambda_0 g_m \gamma^m ). 
\end{align}

Frame fields on $VM$ take the form 
\begin{equation}
  \tilde{e}^M_{\ \, \underline{A}} = 
  \left(\begin{array}{@{}c|c@{}}
    e^m_{\ \underline{a}} & 0 \\\hline
    - \lambda e^p_{\ \underline{a}} g_p & \frac{\lambda}{\lambda_0} 
  \end{array} \right)^M_{\ \, \underline{A}}, \qquad \quad \tilde{e}_{\ \ M}^{\underline{A}} = 
  \left(\begin{array}{@{}c|c@{}}
    e_{\ m}^{\underline{a}} & 0 \\ \hline
    \lambda_0 g_m & \frac{\lambda_0}{\lambda}
  \end{array} \right)^{\underline{A}}_{\ \ M} \, ,
\end{equation}
which are used to define the TW spin connection  
\begin{align} 
    \tilde{\omega}_{M \underline{A} \underline{B}} & = \tilde{e}_{\ \ N}^{\underline{C}} \eta_{\underline{A} \underline{C} } \left( \partial_M \tilde{e}^N_{\ \underline{B}} + \tilde{\Gamma}^N_{\ MP} \tilde{e}^P_{\ \underline{B}} \right).  
\end{align}
The frame fields also allow us to exchanges indices between the flattened Clifford algebras and the spacetime Clifford algebras. Using the spin connection we construct the spinor connection 
\begin{equation} 
    \tilde{\Omega}_B = \tfrac{1}{d+1} \tilde{\omega}_{B[\underline{E}\underline{C}]}\tilde{\gamma}^{\underline{E}}\tilde{\gamma}^{\underline{C}} + V_B, 
\end{equation}
where the choice of $V_B$ in the Roberts gauge gives the components as 
\begin{align} 
    \tilde{\Omega}_b & = \tfrac{1}{d+1} \tilde{\omega}_{b[\underline{E}\underline{C}]}\tilde{\gamma}^{\underline{E}} \tilde{\gamma}^{\underline{C}} - \weight g_b,  \qquad \tilde{\Omega}_{\lambda} = \tfrac{1}{d+1} \tilde{\omega}_{\lambda[\underline{E}\underline{C}]}\tilde{\gamma}^{\underline{E}}\tilde{\gamma}^{\underline{C}}. 
\end{align}

The spin connection induces a Dirac operator, given by 
\begin{align}
    \tilde{\nabla}_M & = \partial_M + \tilde{\Omega}_M, \qquad 
   \tilde{\slashed{\nabla}} = \tilde{\gamma}^M D_M. 
\end{align}
Expanding the implicit sums, the Dirac equation 
becomes 
\begin{align}
    \left( i \slashed{\partial} + \tfrac{i}{2(d+1)} \left( \left( \slashed{\partial} e^n_{\ \underline{b}} \right) \gamma_{n}^{\  \underline{b}} + \mathcal{O} - 2 i \left( \chi + \tfrac{d - \mathfrak{u}}{\lambda_0} \right) \gamma^{\underline{d+1}} \right) + i \tilde{\gamma}^{\lambda} \Xi_{\lambda} - m \right) \Psi = 0, 
\end{align}
where 
\begin{align}
    \mathcal{O} & = \gamma^m \left( \Pi_{nmp} - g_p \left( g_{nm} - \mathfrak{u}_{nm} \right) \right) \left( \gamma^n \gamma^p - \gamma^p \gamma^n \right) - \weight \gamma^m g_m,  \nonumber \\ 
    \chi & = \lambda_0 \left( g_n \Pi^n_{\ mp} - g_p \left( g_m - g_n \mathfrak{u}^n_{\ m} - \mathfrak{a}_m \right) + \partial_m g_p - \mathcal{D}_{mp} + \mathcal{V}_{mp} \right) \gamma^m \gamma^p, \quad \text{and} \nonumber \\ 
    \Xi_{\lambda} & = \partial_{\lambda} - \tfrac{\lambda_0}{\lambda (d+1)} \left( g_n \mathfrak{u}^n_{\ p} + \mathfrak{a}_p \right) \gamma^p \gamma^{\underline{d+1}}.
\end{align}

The Gauss-Bonnet operator is defined as  
\begin{align}
    \mathscr{G}_{AE}^{\ \ \ BFCGDH} & = G_{AE} G^{BF} G^{CG} G^{DH} - 4 \delta^A_{\ C} \delta^E_{\ G} G^{BF} G^{DH} + \delta^A_{\ C} \delta^E_{\ G} G^{BD} G^{FH}.
\end{align}
The Projective Einstein-Hilbert action is 
\begin{equation} 
    S_{\text{PEH}} = \frac{1}{2 \kappa_0} \int \sqrt{|G|} \mathcal{K} d \lambda d^d x,  
\end{equation}
while the Projective Gauss-Bonnet action is 
\begin{equation} 
    S_{\text{PGB}} = J_0 c \int \sqrt{|G|} \left( \mathscr{G}_{AE}^{\ \ \ BFCGDH} \mathcal{K}^{A}_{\ BCD} \mathcal{K}^{E}_{\ FGH} \right) d \lambda d^d x. 
\end{equation}
With both of these terms in place, the TW action is given by 
\begin{align}
    S_{\text{TW}} = S_{\text{PEH}} + S_{\text{PGB}}. 
\end{align}

Minimal coupling follows the standard procedure, up to minor caveats associated with the geometry of $VM$, as described in section \ref{sec: YM Stuff}. The covariant derivative is promoted to a gauge covariant derivative by taking 
\begin{align}
    \tilde{\nabla}_M \to \tilde{\nabla}_M - i \mathscr{A}_M. 
\end{align}
Likewise, the Dirac operator is minimally coupled through the process 
\begin{align}
    \tilde{\slashed{\nabla}} \to \slashed{\mathscr{D}} = \tilde{\slashed{\nabla}} - i \slashed{\mathscr{A}}.
\end{align}

The Projective Riemann tensor with minimal coupling is 
\begin{align}
    \mathscr{K}^A_{\ \ BCD} = \mathcal{K}^A_{\ BCD} - i \delta^A_{\ B} \left( \tilde{\nabla}_{[C} \mathscr{A}_{D]} - i \mathscr{A}_{[C} \mathscr{A}_{D]} \right). 
\end{align} 

The total action involving all of the constituents constructed in this paper is the sum
\begin{align}
    S = S_{\text{TW}} + S_{\text{YM}} + S_{\text{Dirac}},
\end{align}
where the respective components are 
\begin{align}
    S_{\text{TW}} & = \frac{1}{2 \kappa_0} \int \sqrt{|G|} \, \mathscr{K} d \lambda d^d x + J_0 c \int \sqrt{|G|} \mathscr{G}_{AE}^{\ \ \ BFCGDH} \mathscr{K}^{A}_{\ BCD} \mathscr{K}_{A}^{\ BCD} d \lambda d^d x, \nonumber \\ 
    S_{\text{YM}} & = \frac{1}{e^2} \int \sqrt{|G|} \, \mathscr{F}_{AB} \mathscr{F}^{AB} d \lambda d^d x, \quad \text{and} \nonumber \\ 
    S_{\text{Dirac}} & = \int \sqrt{|G|} \, \overline{\Psi} \left( i \slashed{\mathscr{D}} - m \right) \Psi d \lambda d^d x. 
\end{align}

\subsection{Helpful Identities}

The Jacobian and its inverse for a coordinate transformation $X^A \to Y^A$ on $VM$ are 
\begin{align} 
\label{eq: Jacobian on VM}
J^A_{\ B} = \left( \begin{array}{@{}c|c@{}}
    J^a_{\ b} & 0 \\ \hline
    \lambda |J|^{- \frac{1}{d+1}} j_b & |J|^{- \frac{1}{d+1}}
\end{array} \right)^A_{\ \ B} \, , \qquad \bar{J}^A_{\ B} = \left( \begin{array}{@{}c|c@{}}
    \bar{J}^a_{\ b} & 0 \\ \hline
    - \lambda \bar{J}^e_{\ b} j_e & |J|^{\frac{1}{d+1}}
  \end{array} \right)^A_{\ \ B} \, .  
\end{align}
The coordinate transformation for the trace of $\Gamma$ is 
\begin{align}
\label{eq: Trace of Gamma transformation}
    \delta^a_{\ b} \Gamma^e_{\ ec} & \to \delta^a_{\ b} \left( \Gamma^e_{\ em} \bar{J}^m_{\ \ c} - \bar{J}^m_{\ \ c} \partial_m \log |J| \right), \nonumber \\ 
    \delta^a_{\ c} \Gamma^e_{\ be} & \to \delta^a_{\ c} \left( \Gamma^e_{\ me} \bar{J}^m_{\ \ b} - \bar{J}^m_{\ \ b} \partial_m \log |J| \right). 
\end{align}

The curved and flat chiral gamma matrices are related by the identity 
\begin{align}
    \gamma^{\underline{d+1}} = i \gamma^{d+1}. 
\end{align}

\section{Curvature in Palatini Formalism}

\subsection{Curvature in Metric-Affine Gravity}

Let $V^A$ be a vector and $\tau \in [0,1]$ an affine parameter for a loop $\gamma^E$ on $VM$. For equational brevity we write $\gamma_0 = \gamma(0) = \gamma(1)$ as the base point, and use $\dot{\gamma}^E = \frac{d \gamma^E}{d \tau}, \dot{V}^A = \frac{d V^A}{d \tau}$ interchangeably. The change in $V^A$ from transporting along $\gamma$ is   
\begin{align}
\label{eq: Delta V around loop}
    \Delta V^A & = \oint \frac{d V^A}{d \tau} d \tau, 
\end{align}
where $\oint$ should be understood as the integral along $\gamma$. For curvature we are interested in choosing $\gamma$ to be an infinitesimal loop, so that $\delta V^A$ gives the curvature at $\gamma_0$. Because we are parallel transporting $V^A$ we have by definition 
\begin{align}
\label{eq: Basic Parallel Transport Equation}
    \dot \gamma^B \tilde{\nabla}_B V^A = 0. 
\end{align}
Expanding equation \ref{eq: Basic Parallel Transport Equation} with connection coefficients gives 
\begin{align}
    \dot{\gamma}^B \tilde{\nabla}_B V^A & = \dot{V}^A + \tilde{\Gamma}^A_{\ BC} \dot{\gamma}^B V^C = 0, 
\end{align}
from which it follows
\begin{align}
\label{eq: dVdTau negative Gamma}
    \dot{V}^A = - \tilde{\Gamma}^A_{\ BC} \dot{\gamma}^B V^C. 
\end{align}
Substituting equation \ref{eq: dVdTau negative Gamma} into our integral in equation \ref{eq: Delta V around loop}, we have 
\begin{align}
\label{eq: Delta V Integral with negative Gamma}
    \Delta V^A & = - \oint \tilde{\Gamma}^A_{\ BC} \dot{\gamma}^B V^C d \tau. 
\end{align}

It is worth remarking our discussion thus far is valid for any (not necessarily infinitesimal) loop $\gamma$. The necessity of $\gamma$ being infinitesimal now arises: we Taylor expand in $\gamma$ and require second order terms to be negligible. The Taylor expansion around $\gamma_0$ to first order is 
\begin{align}
\label{eq: First Taylor Expansion}
    \tilde{\Gamma}^A_{\ BC} (\gamma (\tau)) & = \tilde{\Gamma}^A_{\ BC} (\gamma_0) + ( \gamma^E - \gamma_0^E ) \partial_E \tilde{\Gamma}^A_{\ BC} (\gamma_0) + \mathcal{O} \big( ( \gamma - \gamma_0 )^2 \big), \nonumber \\ 
    V^A (\gamma(\tau)) & = V^A (\gamma_0) + ( \gamma^E - \gamma_0^E ) \frac{\partial V^A (\gamma_0)}{\partial \gamma^E} + \mathcal{O} \big( ( \gamma - \gamma_0 )^2 \big). 
\end{align}
Here $\partial_E = \frac{\partial}{\partial \gamma^E}$, and the sole reason for expressing this differently between the first and second line in equation \ref{eq: First Taylor Expansion} is later convenience. A different form of the expansion for $V^A$ will be more helpful; chain ruling and another use of equation \ref{eq: dVdTau negative Gamma} changes the expansion for $V^A$ to 
\begin{align}
    V^A (\gamma(\tau)) & = V^A (\gamma_0) + ( \gamma^E - \gamma_0^E ) \frac{\partial \tau}{\partial \gamma^E} \dot{V}^A (\gamma_0) + \mathcal{O} \big( ( \gamma - \gamma_0 )^2 \big) \nonumber \\ 
        \qquad & = V^A (\gamma_0) - ( \gamma^E - \gamma_0^E ) \frac{\partial \tau}{\partial \gamma^E} \big( \tilde{\Gamma}^A_{\ BC} (\gamma_0) \frac{d \gamma^B}{d \tau} V^C (\gamma_0) \big) + \mathcal{O} \big( ( \gamma - \gamma_0 )^2 \big) \nonumber \\ 
        \qquad & = V^A (\gamma_0) - \tilde{\Gamma}^A_{\ BC} (\gamma_0) ( \gamma^B - \gamma_0^B ) V^C (\gamma_0) + \mathcal{O} \big( ( \gamma - \gamma_0 )^2 \big). 
\end{align}
Therefore the Taylor expansion rules we really want are 
\begin{align}
\label{eq: Revised Taylor Expansion}
    \tilde{\Gamma}^A_{\ BC} (\gamma (\tau)) & = \tilde{\Gamma}^A_{\ BC} (\gamma_0) + ( \gamma^E - \gamma_0^E ) \partial_E \tilde{\Gamma}^A_{\ BC} (\gamma_0) + \mathcal{O} \big( ( \gamma - \gamma_0 )^2 \big), \nonumber \\ 
    V^A (\gamma(\tau)) & = V^A (\gamma_0) - \tilde{\Gamma}^A_{\ BC} (\gamma_0) ( \gamma^B - \gamma_0^B ) V^C (\gamma_0) + \mathcal{O} \big( ( \gamma - \gamma_0 )^2 \big). 
\end{align}

Now we are ready to find the curvature. The Taylor expansion of equation \ref{eq: Delta V Integral with negative Gamma} to first order in $\gamma$ is  
\begin{align}
\label{eq: Kurvature Integral to First Order}
    \Delta V^A & = - \oint \left( \tilde{\Gamma}^A_{\ BC} (\gamma_0) V^C (\gamma_0) \right) \dot{\gamma}^B d \tau + \oint \left( \tilde{\Gamma}^A_{\ BC} (\gamma_0) \tilde{\Gamma}^C_{\ EF} (\gamma_0) \gamma_0^E V^F (\gamma_0) \right) \dot{\gamma}^B d \tau \nonumber \\
        & \qquad - \oint \Big( - \tilde{\Gamma}^A_{\ CE} (\gamma_0) \tilde{\Gamma}^E_{\ DB} (\gamma_0) \gamma^D V^B (\gamma_0) + \gamma^D V^B (\gamma_0) \partial_D \tilde{\Gamma}^A_{\ CB} (\gamma_0) \Big) \dot{\gamma}^C d \tau. 
\end{align}
The first term above is
\begin{align}
    - \oint \left( \tilde{\Gamma}^A_{\ BC} (\gamma_0) V^C (\gamma_0) \right) \dot{\gamma}^B d \tau = - \tilde{\Gamma}^A_{\ BC} (\gamma_0) V^C (\gamma_0) \oint \dot{\gamma}^B d \tau = 0,
\end{align}
while the second term is 
\begin{align}
    \oint \tilde{\Gamma}^A_{\ BC} (\gamma_0) \tilde{\Gamma}^C_{\ EF} (\gamma_0) \gamma_0^E V^F (\gamma_0) \dot{\gamma}^B d \tau = \tilde{\Gamma}^A_{\ BC} (\gamma_0) \tilde{\Gamma}^C_{\ EF} (\gamma_0) \gamma_0^E V^F (\gamma_0) \oint \dot{\gamma}^B d \tau = 0. 
\end{align}
Substituting these identities into equation \ref{eq: Kurvature Integral to First Order}, the integral becomes  
\begin{align}
    \Delta V^A & = \oint \gamma^D V^B (\gamma_0) \left( \tilde{\Gamma}^A_{\ CE} (\gamma_0) \tilde{\Gamma}^E_{\ DB} (\gamma_0) - \partial_D \tilde{\Gamma}^A_{\ CB} (\gamma_0) \right) \dot{\gamma}^C d \tau. 
\end{align}
Already there are vivid apparitions of curvature, but half is still unaccounted for. To find the rest of the usual terms, observe integration by parts tells us 
\begin{align}
    \oint \gamma^D \dot{\gamma}^C d \tau = - \oint \gamma^C \dot{\gamma}^D d \tau. 
\end{align}
Therefore $\Delta V^A$ really needs to have an antisymmetrization on these indices. Following this through, we finally have 
\begin{align}
    \Delta V^A & = \frac{1}{2} \oint \gamma^D V^B (\gamma_0) \left( \tilde{\Gamma}^A_{\ [C|E|} (\gamma_0) \tilde{\Gamma}^E_{\ D]B} (\gamma_0) + \partial_{[C} \tilde{\Gamma}^A_{\ D]B} (\gamma_0) \right) \dot{\gamma}^C d \tau \nonumber \\  
    & = \frac{1}{2} \oint \gamma^D \mathcal{K}^A_{\ BCD} (\gamma_0) V^B (\gamma_0) \dot{\gamma}^C d \tau. 
\end{align}
Therefore 
\begin{align}
\label{eq: Definition of Kurvature in Appendix}
    \mathcal{K}^A_{\ BCD} & = \partial_C \tilde{\Gamma}^A_{\ DB} - \partial_D \tilde{\Gamma}^A_{\ CB} + \tilde{\Gamma}^A_{\ CE} \tilde{\Gamma}^E_{\ DB} - \tilde{\Gamma}^A_{\ DE} \tilde{\Gamma}^E_{\ CB}.
\end{align}
This tells us that in the presence of torsion the commutator of covariant derivatives is not the correction definition of $\mathcal{K}$. To reproduce equation \ref{eq: Definition of Kurvature in Appendix} we must therefore take
\begin{align}
    \mathcal{K}^A_{\ BCD} V^B & = [\tilde{\nabla}_C, \tilde{\nabla}_D] V^A + ( \tilde{\Gamma}^B_{\ CD} - \tilde{\Gamma}^B_{\ DC} ) \tilde{\nabla}_B V^A. 
\end{align}

\subsection{Curvature in the Presence of a Yang-Mills Field}

Let all conventions be the same as the previous section. We now want to answer the question of what happens to $V^A$ along $\gamma^E (\tau)$ in the presence of a Yang-Mills field. The computations will be somewhat abbreviated as they are almost identical to the prior section; we believe the detail above justifies any omissions. The parallel transport equation for a minimally coupled Yang-Mills field is  
\begin{align}
    \dot \gamma^B \mathscr{D}_B V^A = 0, 
\end{align}
from which the expansion in components implies 
\begin{align}
    & \dot{V}^A = - \tilde{\Gamma}^A_{\ BC} \dot{\gamma}^B V^C + i \mathcal{A}_B \dot{\gamma}^B V^A. 
\end{align}
Therefore  
\begin{align}
    \Delta V^A & = - \oint \tilde{\Gamma}^A_{\ BC} \dot{\gamma}^B V^C - i  \mathcal{A}_B \dot{\gamma}^B V^Ad \tau.  
\end{align}
As before, is it time to Taylor expand, 
\begin{align}
    \tilde{\Gamma}^A_{\ BC} (\gamma) & = \tilde{\Gamma}^A_{\ BC} (\gamma_0) + (\gamma^E - \gamma_0^E) \partial_E \tilde{\Gamma}^A_{\ BC} (\gamma_0) + \mathcal{O} \left( (\gamma - \gamma_0 )^2 \right), \nonumber \\ 
    V^A (\gamma) & = V^A (\gamma_0) - (\gamma^B - \gamma_0^B) \tilde{\Gamma}^A_{\ BC} (\gamma_0) V^C (\gamma_0) \nonumber \\ 
        & \qquad + i (\gamma^B - \gamma_0^B) \mathcal{A}_B (\gamma_0) V^A (\gamma_0) + \mathcal{O} \left( (\gamma - \gamma_0 )^2 \right), \nonumber \\ 
    \mathcal{A}_B (\gamma) & = \mathcal{A}_B (\gamma_0) + (\gamma^E - \gamma_0^E) \partial_E \mathcal{A}_B (\gamma_0) + \mathcal{O} \left( (\gamma - \gamma_0 )^2 \right). 
\end{align}
On substitution, our integral to first order becomes 
\begin{align}
    \Delta V^A & = \oint \Big( \tilde{\Gamma}^A_{\ BE} (\gamma_0) \tilde{\Gamma}^E_{\ FG} (\gamma_0) - i \tilde{\Gamma}^A_{\ BG} (\gamma_0) \mathcal{A}_F (\gamma_0) - \partial_F \tilde{\Gamma}^A_{\ BG} (\gamma_0) - i \mathcal{A}_B (\gamma_0) \tilde{\Gamma}^A_{\ FG} (\gamma_0) \nonumber \\ 
        & \qquad - \mathcal{A}_B (\gamma_0) \mathcal{A}_F (\gamma_0) \delta^A_{\ G} + i \delta^A_{\ G} \partial_F A_B (\gamma_0) \Big) \gamma^F V^G (\gamma_0) \dot{\gamma}^B d \tau. 
\end{align}
We again must antisymmetrize, giving 
\begin{align}
    \Delta V^A & = \oint \bigg( \partial_{[C} \tilde{\Gamma}^A_{\ D]B} (\gamma_0) + \tilde{\Gamma}^A_{\ [C|E} (\gamma_0) \tilde{\Gamma}^E_{\ D]B} (\gamma_0) \nonumber \\ 
        & \qquad - i \delta^A_{\ B} \left( \partial_{[C} A_{D]} (\gamma_0) - i \mathcal{A}_{[C} (\gamma_0) \mathcal{A}_{D]} (\gamma_0) \right) \bigg) \gamma^D V^B (\gamma_0) \dot{\gamma}^C d \tau, 
\end{align}
or more eloquently, 
\begin{align}
    \Delta V^A & = \oint \left( \mathcal{K}^A_{\ BCD} (\gamma_0) - i \delta^A_{\ B} \mathscr{F}_{CD} (\gamma_0) \right) V^B (\gamma_0) \dot{\gamma}^C \gamma^D d \tau \nonumber \\ 
    & = \oint \mathscr{K}^A_{\ BCD} (\gamma_0) V^B (\gamma_0) \dot{\gamma}^C \gamma^D d \tau.  
\end{align}
Therefore 
\begin{align}
    \mathscr{K}^A_{\ BCD} & = \mathcal{K}^A_{\ BCD} - i \delta^A_{\ B} \mathscr{F}_{CD} \nonumber \\ 
    & = \partial_C \tilde{\Gamma}^A_{\ DB} - \partial_D \tilde{\Gamma}^A_{\ CB} + \tilde{\Gamma}^A_{\ CE} \tilde{\Gamma}^E_{\ DB} - \tilde{\Gamma}^A_{\ DE} \tilde{\Gamma}^E_{\ CB} \nonumber \\ 
        & \qquad - i  \delta^A_{\ B} \big( \partial_C \mathcal{A}_D -  \partial_D \mathcal{A}_C - i  \mathcal{A}_C \mathcal{A}_D + i  \mathcal{A}_D \mathcal{A}_C \big). 
\end{align}
In a similar manner to the previous section, this forces our curvature to be defined as 
\begin{align}
    \mathscr{K}^A_{\ \ BCD} V^B & = [ \mathscr{D}_C, \mathscr{D}_D ] V^A + ( \tilde{\Gamma}^B_{\ CD} - \tilde{\Gamma}^B_{\ DC} ) \mathscr{D}_B V^A. 
\end{align}

\section{The Complete Action}

\label{sec: The Complete Action}

For the sake of illumination we present the expansion of the total action into its constituent fields on $M$. It then takes the form 
\begin{align}
\label{eq: Total Action Expanded}
    S & = \int \sqrt{|G|} \Bigg( \left( \tfrac{1}{e^2} - J_0 c (d-3) \right) F_{ab} F^{ab} - 2 i J_0 c \partial_{[a} \mathfrak{a}_{b]} F^{ab} \nonumber \\ 
        & \qquad + \overline{\psi} W^{-1} \left( i \slashed{\partial} + \slashed{\mathscr{A}} + \tfrac{i}{2(d+1)} \left( \left( \slashed{\partial} e^n_{\ \underline{b}} \right) \gamma_n^{\  \underline{b}} + \mathcal{O} - 2 i \left( \chi + \tfrac{d - \mathfrak{u}}{\lambda_0} \right) \gamma^{\underline{d+1}} \right) + \tfrac{i}{\lambda} \tilde{\gamma}^{\lambda} \Xi - m \right) W \psi \nonumber \\ 
        & \qquad + \tfrac{1}{2 \kappa_0} \mathcal{K} + J_0 c \left(  \mathcal{K}^A_{\ BCD} \mathcal{K}_A^{\ BCD} - 4 \mathcal{K}_{BD} \mathcal{K}^{BD} + \mathcal{K}^2 \right) \Bigg) d \lambda d^d x. 
\end{align}
As before, we have introduced the definitions 
\begin{align}
    \mathcal{O} & = \gamma^m \left( \Pi_{nmp} - g_p \left( g_{nm} - \mathfrak{u}_{nm} \right) \right) \left( \gamma^n \gamma^p - \gamma^p \gamma^n \right) - \weight \gamma^m g_m,  \nonumber \\ 
    \chi & = \lambda_0 \left( g_n \Pi^n_{\ mp} - g_p \left( g_m - g_n \mathfrak{u}^n_{\ m} - \mathfrak{a}_m \right) +  \partial_m g_p - \mathcal{D}_{mp} + \mathcal{V}_{mp} \right) \gamma^m \gamma^p, \quad \text{and} \nonumber \\ 
    \Xi & = \weight - \tfrac{\lambda_0}{(d+1)} \left( g_n \mathfrak{u}^n_{\ p} + \mathfrak{a}_p \right) \gamma^p \gamma^{\underline{d+1}}.
\end{align}

Examining the complete action, the last line is the standard TW action, which has been discussed above. In the first line we have the usual Yang-Mills action from the first term through the factor of $\tfrac{1}{e^2}$. However, we see this is actually modified by a term proportional to $J_0 c$. Again recalling the extremely small value of $J_0$, at low energies this term is negligible, although at very high energy levels this may significantly change the overall coupling strength of any Yang-Mills fields. Considering the second term in this line, we also see the Yang-Mills fields couple to $\mathfrak{a}$, one of the torsion fields. In light of the discussion in section \ref{sec: YM Stuff} this is not surprising, but it is worth noting that this term will allow interactions between Yang-Mills and torsion fields at high energies. Finally, considering the second line involving the expansion of the TW-Dirac action, we see that as before is contains the usual Dirac action in curved spacetime, as well as new contributions to mass and chiral mass terms.

\bibliographystyle{apsrev4-1}

\bibliography{TorsionBibliography}
\end{document}